# Physics-guided deep learning framework for predictive modeling of the Reynolds stress anisotropy

Chao Jiang

*School of Civil Engineering, Harbin Institute of Technology, Harbin 150090, China*

## Abstract

Despite a cost-effective option in practical engineering, Reynolds-averaged Navier−Stokes simulations are facing the ever-growing demand for more accurate turbulence models. Recently, emerging machine learning techniques are making a promising impact in turbulence modeling, but in their infancy for widespread industrial adoption. Towards this end, this work proposes a universal, inherently interpretable machine learning framework of turbulence modeling, which mainly consists of two parallel machine-learning-based modules to respectively infer the integrity basds and closure coefficients. At every phase of the model development, both data representing the evolution dynamics of turbulence and domain-knowledge representing prior physical considerations are properly fed and reasonably converted into modeling knowledge. Thus, the developed model is both data- and knowledge-driven. Specifically, a version with pre-constrained integrity bases is provided to demonstrate detailedly how to integrate domain-knowledge, how to design a fair and robust training strategy, and how to evaluate the data-driven model. Plain neural network and residual neural network as the building blocks in each module are compared. Emphases are made on three-fold: (i) a compact input feature parameterizing the newly-proposed turbulent timescale is introduced to release nonunique mappings between conventional input arguments and output Reynolds stress; (ii) the realizability limiter is developed to overcome the under-constrained state of modeled stress; and (iii) an initial attempt to include constraints of fairness and noisy-sensitivity is made in the training procedure. The influences of the training dataset size, activation function, and network hyperparameter on the performance are also investigated. In such endeavors, an invariant, realizable, unbiased, and robust data-driven turbulence model is achieved, and does gain good generalization across channel flows at different Reynolds numbers and duct flows with various aspect ratios.

## I. Introduction

In flows of engineering interest, Reynolds-averaged Navier−Stokes (RANS) simulation keeps an ongoing popular alternative to high-fidelity, but computationally expensive methods, e.g., direct numerical simulation (DNS) and large eddy simulation (LES). Despite significantly reducing the computational complexity of the nonlinear Navier−Stokes system, RANS simulations inevitably introduce considerable inadequacies mainly due to inadequate turbulence models in representing the real turbulence physics, which has been well documented in the literature.[1-3] There is, on the other hand, ever-growing demand for improved RANS models: an ambitious desire to apply a reliable RANS-based prediction over the entire flight envelope in the foreseeable future.[4,5] Even if DNS and LES eventually are feasible in engineering, they are expected to rely mainly on RANS-based near-wall models. The needs are clear, while conventional model developments almost remain stalled, indicating a need for new ideas going forward.

Very recently, encouraging advances are being made in improvements of model accuracy using the cutting-edge machine learning (ML) techniques.[6,7] ML-augmented turbulence modeling with data-driven aspects provides a paradigm shift for turbulence modeling to overcome the well-known bottleneck of conventional intuition-based model creations. The first attempt to apply an ML algorithm to turbulence modeling can go back to Cheung et al.,[8] who developed a Bayesian statistical model for the closure coefficients of Spalart-Allmaras model. Since then, data-driven turbulence modeling has been growing vigorously. Edeling et al.[9,10] and Zhang et al.[11] performed Bayesian inference on all closure coefficients of linear eddy-viscosity models and accompanying transport equations. Stochastic modeling always makes the obtained coefficients being spatial-independent. Instead, Weatheritt and Sandberg[12-14] use gene expression programming to explicitly recalibrate the closure coefficients of nonlinear algebraic models as functions of the local flow field. Zhu et al.[15] reconstructed a function-surrogated model to approximate the turbulent viscosity using the radial basis function. It is worth noting that, Ling et al.[16] and Jiang et al.[3] pioneered the use of deep neural networks (DNNs) to gain spatially-varying coefficients of nonlinear algebraic models. All aforementioned efforts are part of ML-augmented inverse modeling of the closure coefficients given a form-constraint RANS model.

Rather than recalibrating a set of coefficients to reduce the parametric uncertainty, adding corrective source terms is another route to correct the existing RANS models, which focuses on improving the structural uncertainty. Dow and Wang[17] made the first attempt to statistically estimate the discrepancy between the inferred and modeled turbulent viscosity of the $k-\omega$ model using a Gaussian process. To achieve a dependence on the feature space, Duraisamy and co-workers[18-20] proposed a promising data-driven framework comprised of Bayesian-based field inversion and ML-based function reconstruction to build the mapping of inferred discrepancy (e.g., a corrective multiplier or a direct corrective source) to local flow quantities. Following this practice, He et al.[21] adopted continuous adjoint formulation to correct the Spalart-Allmaras model, and Yang et al.[22] combined Bayesian and DNN to correct the $k-\omega-\gamma-A_r$ transition model. Iaccarino and co-workers[23-27] presented a novel approach to account for structural uncertainty by directly perturbing the elements in the spectral decomposition of Reynolds stress. Although not an ML algorithm, this approach has two prominent advantages: (i) good interpretation because of turbulent kinetic energy, eigenvalue, and eigenvector representing the magnitude, shape, and orientation of perturbations; and (ii) easy implementation of realizability. Xiao and co-workers[28-30] benefited from both above-mentioned practices and successfully used random forest to predict the discrepancy in the Reynolds stress projections. Yin et al.[31] followed this direction using DNN.

In comparison, the greatest challenge in the philosophy of turbulence modeling is to develop a universal model across multiple classes of flows (the ultimate barrier). This requires a more flexible framework to reduce both structural and parametric uncertainties, aiming at developing new models free from the existing RANS models, rather than improving them. Early attempts have been made to directly predict eigenvalues of stress anisotropy using ML algorithms,[32,33] which can only account for structural uncertainty. It should be noted that ML-augmented data-driven approaches have successfully found hidden exact relations behind data;[34-37] however a universally accurate turbulence model cannot be waiting to be uncovered because of loss information in the closure terms. Alternatively, the optimal models in a user-defined sense may be available.

In a similar direction, we design two parallel ML-based modules to infer both integrity bases and closure coefficients based on data and domain-knowledge, which reduce structural and parametric uncertainties, respectively. The two modules are interactive through a multiplicative layer by analogy to tensor modeling, thus forming an interpretable framework of turbulence modeling. Each learned integrity basis with corresponding referred closure coefficient makes up one term of the overall polynomial to construct a functional model-from. There has been a strong belief that different flows could be simulated properly using the same form-constraint model but with possible different coefficients, indicating that closure coefficients reflect the physical effects from the flow boundary. As a result, the proposed framework can serve as a transfer learning framework for dynamic tasks: one can retrain the data-driven model by only re-updating the closure coefficients with fixed learned integrity bases (as learned modeling knowledge). It is an advantage when compared to models using a single ML-based module to directly predict Reynolds stress. Given the universal approximation, a DNN (i.e., deep learning) is selected as the machine learning algorithm, which has achieved great success in many fields, such as image recognition and natural language processing. Kutz[38] stressed the major advantages of DNN in extracting features from informative data and capturing invariances. Ling et al.[39] further demonstrated that DNN has a clear advantage of learning invariances through data augmentation over random forests.

To investigate the performance of the proposed framework by itself, a controlled environment will be needed in which the learning dynamics of the framework, e.g., sources of gains and the upper-performance limit, can be explored in isolation. Towards this goal, we first constrain the integrity bases to conventional tensor bases proposed by Lumley[40] and Pope[41]. Against this backdrop, we conduct this work and perform a systematic investigation on following three-fold: (i) how to select input features and utilize prior domain-knowledge; (ii) how to set up a fair and robust learning strategy to overcome or moderate prediction prejudice and noisy-sensitivity; and (iii) how to evaluate a data-driven model. We also note that such a version with pre-constrained integrity bases still represents a broad class of data-driven turbulence modeling and previous many works (e.g., Jiang et al.,[3] Weatheritt and Sandberg,[13] and Ling et al.[16]) are its special cases. It could be expected that the class of methodology is very useful both in improving the existing models and in assisting to develop new models. Thus, it is naturally a logical starting point to explore a controlled version.

Overall, this work seeks to develop a universally interpretable machine learning (UIML) framework of turbulence modeling with build-in fundamental principles, under which a physics-informed residual network (PiResNet) is achieved. The remainder of the paper is structured as follows. Section II introduces some fundamental aspects of UIML, focusing especially on integrating structural domain-knowledge into PiResNet, e.g., feature selection as inputs, physical constraints of outputs and an ML architecture by analogy to tensor modeling. Section III covers a fair-awareness and robust training strategy for PiResNet, aiming at achieving a high-fidelity representation of turbulence closures without imbalance-induced generalization prejudice and lower noisy-sensitivity. Section IV concerns PiResNet performance on prediction stages, including extrapolative accuracy, realizability and robustness. Finally, Section V draws the main findings and highlights some remaining key challenges for future extensions of this work.

## II. Methodology

For incompressible turbulent flows, the RANS momentum equations can be stated as

$$\frac{\partial \bar{\boldsymbol{u}}}{\partial t} + \bar{\boldsymbol{u}} \cdot \nabla \bar{\boldsymbol{u}} + \nabla \tilde{p} - \nu \nabla^2 \bar{\boldsymbol{u}} = -2\nabla \cdot k\boldsymbol{a}, \tag{1}$$

where $\bar{\boldsymbol{u}}$, $\tilde{p}$, $k$, and $\nu$ are the mean-flow velocity, nominal pressure (including kinematic pressure and isotropic part of Reynolds stress), turbulent kinetic energy, and kinematic viscosity, respectively. The stress anisotropy tensor $\boldsymbol{a} \equiv \boldsymbol{\tau}/(2k) - \mathbf{I}/3$ where $\boldsymbol{\tau}$ and $\mathbf{I}$ are respectively the Reynolds stress and identity tensors, represents effects of unresolved turbulence and needs closure modeling. The stress anisotropy can only be related to the mean-flow field.

Despite some progress[3,42-45] in recent years, efforts in conventional turbulence modeling have considerably diminished. There may be the following reasons. It is very difficult to derive, in theory, an accurate analytic model only based on the developer's limited knowledge, experience, and intuition. Even if a more accurate model-form is achieved, its potential benefits still depend on a set of coefficients to be determined. Jiang et al.[3] have demonstrated in detail that conventional methods to calibrate the closure coefficients are facing many disadvantages. The existing rich data from high-resolution simulations and experiments, on the other hand, is creating favorable conditions to inform closure models. ML algorithms have a powerful capability in distilling more informative knowledge from raw data. Thus, the present work is motivated to develop an ML-augmented framework by fusion of data and domain-knowledge to reconstruct a mapping from the mean-flow field to the stress anisotropy.

### A. Summary of the proposed UIML framework

Turbulence modeling, at the RANS level, is a classic example of multivariate problems where the stress anisotropy tensor must be modeled to represent nonlinear interactions between mean-flow and turbulent motion in different directions. Tensor modeling is well known as a good physical option in multivariate problems due to its objectivity. Generally, a given multivariate problem can be formulized to a polynomial of a set of integrity bases with corresponding closure coefficients, just like Lumley[40] and Pope[41] did. Accordingly, we design two parallel ML-based modules, as depicted in Fig. 1, to be responsible for distilling integrity bases and closure coefficients from high-fidelity data, both of which are integrated through a multiplicative layer to constitute the stress anisotropy tensor. In either module, there is enough flexibility to build the desired DNN architecture, e.g., plain neural network, residual neural network[46,47] and Bayesian neural network,[48-50] thus indicating a universal, inherently interpretable ML-based framework for turbulence modeling (denoted as UIML). According to the well-defined concept by Rudin,[51] the interpretability of the resulting UIML, first and foremost, lies in its constrained model-form as

$$f_1: \boldsymbol{q} \mapsto \boldsymbol{g}, \quad f_2: \mathcal{Q} \mapsto \mathcal{T}, \quad \boldsymbol{a} = \boldsymbol{g} \cdot \mathcal{T}, \tag{2}$$

where $\boldsymbol{q} = \{q_1, q_2, \cdots\}$ and $\boldsymbol{\mathcal{Q}} = \{\boldsymbol{Q}^{(1)}, \boldsymbol{Q}^{(2)}, \cdots\}$ are two vectors of scalar-valued input features $q_i$ and tensor-valued input features $\boldsymbol{Q}^{(i)}$; $\boldsymbol{g} = \{g_1, g_2, \cdots\}$ and $\boldsymbol{\mathcal{T}} = \{\boldsymbol{T}^{(1)}, \boldsymbol{T}^{(2)}, \cdots\}$ are two vectors of learned scalar-valued closure coefficients $g_i$ and tensor-valued structural bases $\boldsymbol{T}^{(i)}$. The resulting model represented by Eq. (2) has the structure $\boldsymbol{a} = \sum g_i \boldsymbol{T}^{(i)}$, which helps us to interpret what the UIML has learned from informative data. $f_1$ and $f_2$ account for parametric and structural representations of turbulence physics, respectively. The modeling process is straightforward. The transparency of the UIML obtains the superiority of symbolic regression based on evolution algorithms.[12,13] In contrast, Edeling *et al.*[52] stressed that there is no analytic form for interpreting the structure of learned model discrepancies by other data-driven approaches.[18,22,29] In particular, when the structural bases $\{\boldsymbol{T}^{(i)}\}_{i=1}^{10}$ are constrained to those of Pope,[41] the UIML degrades to a general form of conventional algebraic models. Thus, the UIML is a more universal framework than existing models based on representation theorems (e.g., Hilbert basis theorem[53] and Cayley-Hamilton theorem[54]), and has more flexibility. Second, the UIML obeys *prior* domain knowledge that we inject into it hereafter, such as invariant properties (rotational, reflectional, extended Galilean, and scale invariances), corrective turbulent timescale, realizability constraints, unbiased principles, and robust regularizations. Alber *et al.*[55] conducted a systematic investigation of ML-based multiscale modeling with no fundamental laws of physics, and encountered unphysical solutions and ill-posed problems. Chang *et al.*[56] and Pawar *et al.*[57] also stressed the significance of domain knowledge. All *prior* domain knowledge is thus reasonably converted into modeling knowledge for the UIML during our model development. Third, the unbiased and robust learning strategy adopted in this work helps interpret how the UIML achieves fair and noise-insensitive predictions. Developing an inherently interpretable model is of critical importance for physical problems. However, interpretability has not been reasonably considered. The UIML is an attempt to bridge this gap.

After designing UIML, we first utilize training data to establish a data-driven surrogate model of the stress anisotropy by minimizing the following objective-function to determine the optimal trainable parameters $\boldsymbol{\theta}_\pi$ of the adopted ML algorithm

$$\boldsymbol{\theta}_\pi = \arg\min_{\boldsymbol{\theta}} \{ \|\tilde{\boldsymbol{a}} - \boldsymbol{g} \cdot \boldsymbol{\mathcal{T}}\|_F^2 + \lambda_g \|\boldsymbol{g}\|_1 + \lambda_w \|\boldsymbol{w}\|_2^2 \}, \quad (3)$$

where $\|\cdot\|_F$, $\|\cdot\|_1$, and $\|\cdot\|_2$ denote Frobenius norm, $L$-1 norm, and $L$-2 norm, respectively. $\tilde{\boldsymbol{a}}$ is targeted stress anisotropy from high-fidelity data. The first term in the objective-function is the training error representing the quality of training. The second term is a sparsity-inducing penalty on the closure coefficients, and the third term is noisy-insensitive (i.e., robust) penalty on the trainable parameters of the ML algorithm. $\lambda_g$ and $\lambda_w$ are penalty factors are penalty factors and trainable parameters are updated using the mature backpropagation algorithm.[58] More details in Section III. The resulting data-driven model reads

$$\boldsymbol{a} = f(\boldsymbol{q}, \boldsymbol{\mathcal{Q}}; \boldsymbol{\theta}_\pi). \quad (4)$$

Then we insert the learned model into the computational fluid dynamics (CFD) environment represented by Eq. (1) and evaluate its predictive performance for quantities of interest (QoIs) in unseen flows. The overall procedure is briefly summarized as follows:

(1) Data collection and preparation. Collect representative datasets for training and testing flows from high-resolution experiments and high-fidelity simulations, and reformat all raw data with a unified standard.

(2) Feature selection and computation. Select input features reasonably according to the general principles (e.g., mean-flow field variables, invariance properties, compactness) and compute input features and targeted stress anisotropy both for training and testing flows.

(3) Data sampling and model training. Develop a fair sampling technique to obtain a training dataset with enough sample diversity and design a fair and robust learning strategy (e.g., unbiased outcomes for different flow domains, stress components and coordinate systems, noisy-insensitivity to input features) to achieve a desired data-driven model.

(4) Prediction and evaluation. Perform the learned model in unseen flows and evaluate its predictive skills (e.g., efficiency, accuracy, robustness).

The overall workflow of UIML is shown in Fig. 1. Prior domain-knowledge such as physically-motivated selection of input features and corrective turbulent timescale considering viscous effects, it should be noted, is considered in the design phase (Step 2). Realizability properties to constrain the learned model are enforced both in the training (Step 3) and testing phases (Step 4). A scale- and frame-invariant objective-function as well as regularization constraints both on trainable parameters and closure coefficients are implemented in Step 3. Finally, the propagation test of modeled stress anisotropy to QoIs in predictive settings is conducted in Step 4. By fusion of informative data and domain-knowledge, the developed model under UIML is both data- and knowledge-driven. Each component is detailed hereafter.

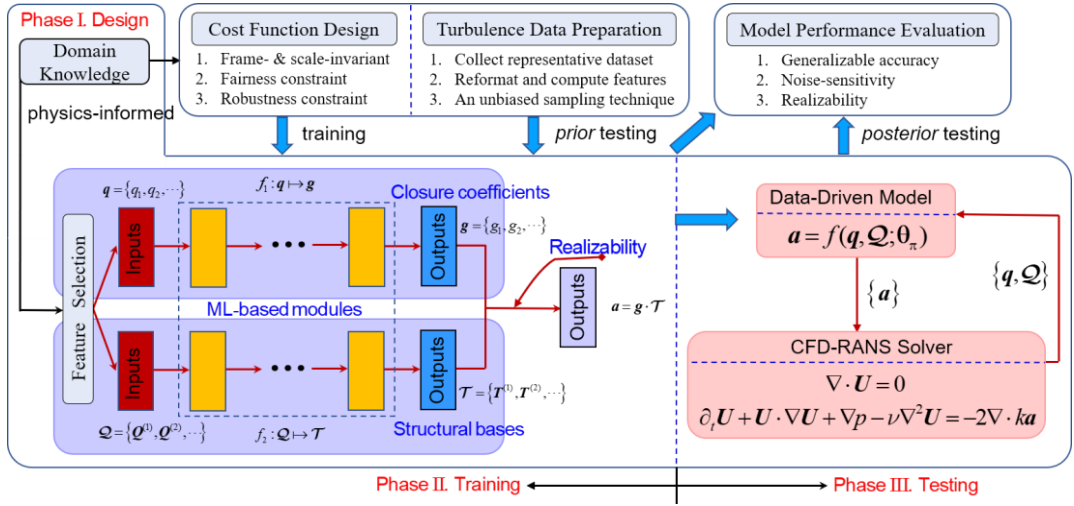

FIG. 1. The development and evaluation lifecycle of the interpretable machine-learning framework (UIML) for turbulence modeling containing three phases: a design of UIML integrated with domain-knowledge (Phase I), a fair and robust training strategy for UIML (Phase II), and a systematic performance testing of UIML (Phase III).

Using the procedure described above, the mappings $f_1: q \mapsto g$ and $f_2: \mathcal{Q} \mapsto \mathcal{T}$ are eventually learned as regression functions. Rather than stochastic functions by Gaussian process[17] or Bayesian inversion[8,11,59], we aims at developing a deterministic predictor using DNNs. ML-based deterministic modeling is computationally cheaper than stochastic modeling. Besides, stochastic modeling always depends on data space when using Gaussian process or on parameter space when using Bayesian inversion, both not on feature space. Such models have no physical content other than the assumption of smoothness. Stochastic models also encounter poor extrapolation capability due to their geometry-dependence. Finally, stochastic models targeted at the mean-flow velocity do not satisfy realizability constraints for stress anisotropy. In comparison, deterministic modeling using DNNs can move beyond these limitations.

In particular, all reasonable principles regarding conventional turbulence modeling are strictly respected in data-driven turbulence modeling under UIML. Most importantly, given rotational and reflectional invariances[60] of RANS equations represented by Eq. (1), the overall regression functions $f$ representing the data-driven model must be form-invariant w.r.t. rotations and reflections of the coordinate system. That is, a transformation of $f$ under arbitrary orthogonal tensor $R$ ($RR^T = R^TR = I$ and $\det R = \pm 1$) should satisfy the relation as

$$f(q, R\mathcal{Q}R^T; \theta_\pi) = R[f(q, \mathcal{Q}; \theta_\pi)]R^T. \tag{5}$$

The superscript T means the transpose of a tensor. It should be noted that RANS equations obey invariances under rotational and reflectional transformations of the coordinate system while none of terms in RANS equations (including the stress anisotropy tensor) are rotational or reflectional invariances. Accordingly, we seek to the rotational or reflectional invariances on regression functions $f$, not on the stress anisotropy itself. Obviously, the scalar-valued functions $f_1 : q \mapsto g$ as functions of all frame-invariant scalars is invariant w.r.t. rotations and reflections of the coordinate system. Thus, Eq. (5) reduces to

$$R\mathcal{T}R^T \equiv Rf_2(\mathcal{Q})R^T = f_2(R\mathcal{Q}R^T). \tag{6}$$

To satisfy Eq. (6), $f_2 : \mathcal{Q} \mapsto \mathcal{T}$ should be designed as tensor-valued isotropic functions of the arguments. Speziale et al.[61] and Pope[41] explicitly constructed such functions for pressure-strain correlation and the stress anisotropy, respectively. In comparison, Eq. (6) herein is represented by an ML-based module. However, it is always misunderstanding that the input-output equivariance can be guaranteed by choosing rotational and reflectional invariants as input features, for instance, that is just the case of Wang et al.[28] and Wu el al.[29] Considering that there is no explicit analytic form (as an aside, strictly existing but considerably complex to understand) in most data-driven turbulence modeling, one must take care of the construction of pre-defined functions by adopted ML algorithms. Ling et al.[39] used data augmentation to preserve approximate invariances under rotational and reflectional transformations. Furthermore, regarding the dynamical processes of RANS equations, the Reynolds stress $\tau$ reduces from Euclidean invariance to extended Galilean invariance,[62] not just Galilean invariance as usually stated. It is therefore easily deduced that the stress anisotropy $a$ is also extended Galilean invariant owing to extended Galilean invariance of both turbulent kinetic energy $k$ and Reynolds stress $\tau$. When using isotropic functions, extended Galilean invariance of the predicted stress anisotropy $a$ can be easily guaranteed by constraining extended Galilean invariance to input features $\{q, \mathcal{Q}\}$. However, input features of most previous works such as Xiao et al.,[28-30] Zhu et al.,[15] Yin et al.,[31] and Ling and Templeton,[63] are not extended Galilean invariant (even some are not Galilean invariant), which could destroy the properties of predicted Reynolds stress and the extrapolation capability. Notably, Ling et al.[16,39] also confused the two concepts of rotational/reflectional invariance and Galilean invariance (an invariance under an inertial frame with constant translations). Besides, dimensionless stress anisotropy is scale-invariant owing to the Reynolds-number similarity of RNAS equations. Nondimensional input features are therefore needed. Lastly, the stress anisotropy is bounded by a set of inequality constraints, i.e., realizability requirements, which have been comprehensively studied by Lumley[64] and Banerjee et al.[65] However, these constraints are not always satisfied using existing limiters, e.g., Ling et al.[16] (shown in Section IV. A). All four considerations mentioned above are important prerequisites of modeling stress anisotropy and presented below. We focus on clarifying some existing unreasonable practices.

B. Data-driven and knowledge-driven aspects

A large amount of available data does provide a special opportunity for ML algorithms to inject new vitality into turbulence modeling. However, this does not mean that existing knowledge about turbulence modeling should be shelved. Alber et al.[55] demonstrated that pure data-driven models (in biological, biomedical and behavioral sciences) always encounter difficulties in achieving reliable and interpretable results. On the one hand, although data contains rich information about a physical system to be modeled, a complete dataset or a complete set of input features is not always available in practice. Complete information thus cannot be fed into the data-driven model. Some redundant and irrelevant input features even play the opposite role. Even if such a complete and compact dataset and input features are eventually achieved, on the other hand, it is still difficult for ML algorithms to directly learn all precise physical relationships, especially inequities and qualitatively physical properties. For instance, pure data-driven models can easily obtain sufficiently accurate predictions for stress anisotropy components (let alone nonzero training error for the extrapolation consideration), but difficult to satisfy the realizability constraints which provides a set of inequalities for stress anisotropy components. Finally, without any prior domain-knowledge, pure data-driven models will pay the price in other ways. Wu et al.[29] and Ling et al.[39] used data augmentation to preserve invariances of regression functions which significantly increased the computational cost. More than that, the pure data-driven methodology is prone to encounter convergence difficulty in training stages and potential failure risk in prediction stages (not intrinsic invariances). It is worth noting that pure data-driven models have been facing ill-posed problems (due to thousands of network parameters to be determined) and non-physical solutions.

Thus, existing domain-knowledge must be reasonably considered in synergy with data-driven aspects to accelerate turbulence modeling. The present work is motivated to develop a data- and knowledge-driven turbulence model. Section II.A outlines what domain-knowledge will be embedded in the data-driven model and how domain-knowledge helps achieve a inherently interpretable model. Addtionally, domain-knowledge introduced in this work helps UIML gain other benifits: (i) siginificant reduction on the need for training data with invariaces, (ii) preserving phsyical solutions with realizability constraints, (iii) improvements on fair predictions with a fair leanring statrgy, and (iv) improvements on robust predictions with regularization of both trainable parameters and closure coefficients. In the following, we are centered about how to intergrate these domain-knowledge into the developed model.

## C. A general principle to select input features

The present work seeks to a deterministic model depending on feature space rather than a spatial-independent stochastic model. Thus we need to construct proper input features from the mean-flow field $\{\nabla \bar{\boldsymbol{u}}, \nabla p, k, \varepsilon, \nu\}$ where $\varepsilon$ is the dissipation rate of turbulent kinetic energy. The fundamental principle to select input features $\{\boldsymbol{q}, \mathcal{Q}\}$ is according to the requirements of targeted stress anisotropy $\boldsymbol{a}$ and the properties of regression functions $f$. It should be noted that the stress anisotropy $\boldsymbol{a}$ rather than Reynolds stress $\boldsymbol{\tau}$ is selected as the predicted target. Two dynamically similar flows share the same $\boldsymbol{a}$ but different $\boldsymbol{\tau}$. Thus, one can benefit from two aspects when using $\boldsymbol{a}$ as the predicted target: (i) a more compact training data to reduce computational cost, and (ii) a scale-invariant objective-function to guarantee the validity of hyper-parameters in ML algorithms across dynamically similar flows.

For targeted stress anisotropy $\boldsymbol{a}$, there are four physical principles: (i) extended Galilean invariance,[62] (ii) scale invariance, (iii) symmetry, and (iv) realizability constraints.[64] To satisfy rotational and reflectional invariance, $f_1$ and $f_2$ are isotropic functions "implicitly" represented by ML-based modules. Thus, the input features must obey the same principles as targeted stress anisotropy $\boldsymbol{a}$ except for the last principle. Realizability constraints are a set of inequalities among stress anisotropy components, thus being considerably difficult to satisfy by constructing input features. Instead, realizability constraints are directly enforced to predicted stress anisotropy $\boldsymbol{a}$. Besides above-mentioned requirements, the selected input features must have clear physical justifications.

The mean strain-rate and rotation-rate tensors $\boldsymbol{S} \equiv (\nabla \bar{\boldsymbol{u}} + \nabla \bar{\boldsymbol{u}}^{\mathrm{T}})/2$ and $\boldsymbol{\Omega} \equiv (\nabla \bar{\boldsymbol{u}} - \nabla \bar{\boldsymbol{u}}^{\mathrm{T}})/2$ are widely used field variables in flow analysis and conventional turbulence modeling. Considering that two training and testing flows are homogenous in the streamwise direction, only $\nabla p$ is excluded, leading to a general mapping from the mean-flow field to the stress anisotropy

$$\boldsymbol{a} = \boldsymbol{a}(\boldsymbol{S}, \boldsymbol{\Omega}, k, \varepsilon, \nu). \tag{7}$$

Arguments of Eq. (7) $\boldsymbol{S}$, $\boldsymbol{\Omega}$, $k$, $\varepsilon$, and $\nu$ are extended Galilean invariant. Regarding extended Galilean invariance, there are two things to be clarified. One is that extended Galilean invariance of input features have been rarely considered reasonably to reproduce extended Galilean invariance of the stress anisotropy. In previous many works such as Zhu et al.,[15] Wang et al.[28] and Ling and Templeton[63], several input features are not Galilean invariant (also not extended Galilean invariant); part of input features by Wu et al.[29,30] are not extended Galilean invariant. Such input features do destroy the physical properties of the stress anisotropy, i.e., qualitatively incorrect. Another is an interesting but somewhat controversial topic about the validity of frame-indifference in turbulence modeling. Earlier works of Speziale and Charles[66] arrived at erroneous conclusions, not considering the dynamical process, that $\boldsymbol{a}$ is frame-indifference – a guiding principle in continuum mechanics put forth by Noll.[67] Consequently, $\boldsymbol{\Omega}$ was excluded in turbulence modeling because it is not frame-indifference. Later, Lumley[40] argued, based on experimental observations, that $\boldsymbol{\Omega}$ should be introduced into the constitutive model to account for effects of rigid rotation on Reynolds stress.

Wu et al.[29] hold the same idea, indicating that $\Omega$ only exists in some special flows. However, Huang and Durst[62] recently demonstrated analytically that $a$ is not frame-indifference, implying that $\Omega$ can serve as an effective constitutive argument in turbulence modeling and not just in some flows. Thus, $a$ is not independent of the observers.

For integrity bases inference $f_2 : \mathcal{Q} \mapsto \mathcal{T}$, input features and outputs should be dimensionless and extended Galilean invariant. Outputs also should be symmetric and realizable. Thus, we select $s \equiv kS/\varepsilon$ and $\omega \equiv k\Omega/\varepsilon$ as input features. Specifically, we constrain regression functions $\mathcal{T} = \{T^{(i)}\}_{i=1}^{10}$ to a general form of algebraic models by Pope[41], thus achieving a special version of UIML as

$$a = g_1 s + g_2 (s^2 - \frac{1}{3}\mathrm{tr}(s^2)\mathbf{I}) + g_3 (s\omega - \omega s) + g_4 (\omega^2 - \frac{1}{3}\mathrm{tr}(\omega^2)\mathbf{I}) + \cdots \qquad (8)$$

where $\mathrm{tr}(\cdot)$ denotes the trace. For coefficients inference $f_1 : q \mapsto g$, we select traditional flow variables (related to the traces) as input features. See more details in Refs.[16,41] For instance, first two terms $s_m \equiv (2s_{kl}s_{kl})^{1/2}$ and $\omega_m \equiv (2\omega_{kl}\omega_{kl})^{1/2}$ are extended Galilean, rotational/reflectional and scale-invariants and related to the positive second invariant of $\nabla\bar{u}$, i.e., $Q_c \equiv (\omega_m^2 - s_m^2)/4$, which is a criterion[68] to detect vortical structures. It is worth noting that Eq. (8) cannot achieve satisfactory predictions only depending on conventional tensor invariants $q = \{s_m, \omega_m, \ldots\}$ and an additional input feature must be introduced. In Section II.D, the reasons of the failure of Eq. (8) are detailly analyzed. Previous works of Ling et al.[16] and Weatheritt and Sandberg[13] encounter this difficulty, which is also noticed recently by Geneva and Zabaras.[69] The present work provides a solution.

Additionally, the requirement that input features must be dimensionless is also from ML algorithms to solve physical problems. It is very difficult for ML algorithms to guarantee the expected dimension of targeted outputs when using dimensional input features due to the uncontrollable and complex nested functional form of ML algorithms. Such a model is qualitatively incorrect even if a good prediction may be provided numerically. This aspect is always ignored.

Finally, input features should be compact and complete for physical problems. Due to the lack of prior knowledge about arguments, it is difficult to obtain ideally complete input features. One possible way in the near future is to perform feature compression and extraction from redundant input features using autoencoder. With more input features, the learned model is more complex and more difficult to understand. For physical problems, models are expected to be as simple as possible which is well-known as "Occam's Razor". The performance of our model is significantly improved by adding only one input feature as shown later.

It cannot be overemphasized that the construction of input features is of critical importance when applying ML algorithms to physical problems. When physical properties of input features are not satisfied, data augmentation as an alternative results in noncompact training data and an increase in computational cost. Ling and Templeton[63] have built a procedure to select input features and Wu et al.[29] have extended it. The present work focuses on the general principles and perfect them.

### D. Preserving uniqueness with a new timescale

To establish a deterministic ML-based predictor (i.e., the data-driven model), a unique mapping from selected input features to targeted outputs in the training dataset is a fundamental prerequisite.

However, using $q = \{s_m, \omega_m, \ldots\}$ and $\mathcal{Q} = \{s, \omega\}$ as input features cannot satisfy this condition in turbulent flows. We perform Eq. (8) in two-dimensional fully-developed channel flows to demonstrate the drawbacks and analyze the causes. In such flows, only $dU/dy$ ($U$ is the mean streamwise velocity and $y$ denotes the wall-normal direction) exists. Accordingly, all nonzero mean-flow quantities are listed below

$$s_{12} = s_{21} = \frac{1}{2} s_m, \quad \omega_{12} = -\omega_{21} = \frac{1}{2} \omega_m, \quad s_m = \omega_m = \frac{k}{\varepsilon} \frac{dU}{dy}. \quad (9)$$

The flow is controlled by the dominant shear stress $a_{12}$. Substituting Eq. (9) into Eq. (8) leads to

$$a_{12} = \frac{1}{2} g_1 s_m - \frac{1}{4} g_6 s_m \omega_m^2. \quad (10)$$

where coefficients $g_1$ and $g_6$ are ultimately dependent on $s_m$. Thus, the resulting Eq. (10) is expressed as a one-variate function of $s_m$. In contrast, Fig. 2(a) gives the true relationship between $a_{12}$ and $s_m$ at $Re_\tau = 5200$,[70] exhibiting unexpected nonunique mappings, i.e., $s_m(y_A^+) = s_m(y_B^+)$, and $a_{12}(s_m)|_{y_A^+} \neq a_{12}(s_m)|_{y_B^+}$. Obviously, Eq. (10) is inconsistent with Fig. 2(a). As a result, Eq. (8) cannot be used as the regression function to establish a functional mapping from selected $s_m$ to targeted $a_{12}$ given a training dataset.

So what will happen when Eq.(8) is trained on this inherently inconsistent data? Data-driven models of Ling et al.[16] (denoted as TBNN) and Weatheritt and Sandberg[13] can also be represented by Eq. (8). The only difference is the truncation order: the former is fully complete while the latter is 3. We take TBNN as an example to quantitatively illustrate the drawbacks of original Eq.(8). We perform two training scenarios: TBNN-1 is trained on the whole data while TBNN-2 is trained using the data far from the wall ($y^+ > 9$). Both results are shown in Fig. 2(b). TBNN-1 shows worse training performance than TBNN-2 (nonunique mappings only in the circle-marked region as shown in Fig. (1)), indicating that nonunique mappings in the training dataset make training more difficult. Additionally, the training process of TBNN-1 shows bias toward the region where more data exists, as shown in Fig.2 (b), because more data means bigger contribution to the overall training error. Obviously, TBNN-1 is going to get worse when the amount of training data is equal in regions far from the wall and near the wall, and at the same time achieves equally bad predictions for both regions. It is noted that this is why we design a fair data sampling technique to achieve fair predictions for different regions in Section III.B. Lastly, both trained TBNN-1 and TBNN-2 provide worse predictions of $a_{12}$ in the same flow where only meshes are different from those in training cases, indicating that TBNN-1 and TBNN-2 merely fit on the existing data and have no generalization capability. Recently, Geneva and Zabaras[69] also noticed that TBNN is difficult to train and yields satisfactory predictions. This problem is common for models based on Eq. (8).

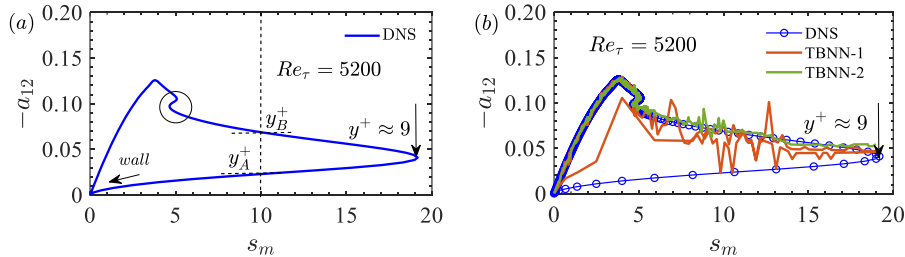

FIG. 2. Two-dimensional fully-developed channel flow at $Re_\tau = 5200$: (a) the shear stress anisotropy $a_{12}$

against characteristic strain rate $s_m$ from DNS,[70] and (b) two TBNN-predicted shear stress anisotropy $a_{12}$ against characteristic strain rate $s_m$ with different meshes from that of DNS.

From the mathematical standpoint, the number of input features is not enough. The relations $s(y_A^+) = s(y_B^+)$ and $\omega(y_A^+) = \omega(y_B^+)$ result in the same values for the structural bases at locations A and B, i.e., $T^{(i)}(y_A^+) = T^{(i)}(y_B^+) (i = 1, 2, \ldots, 10)$. To return different values for $a$ at locations A and B as expected in Fig. 2(a), $g_i(y_A^+) \neq g_i(y_B^+)$ must be guaranteed. However, Eq. (8) always provides the same value $g_i(y_A^+) = g_i(y_B^+)$. Obviously, Eq. (8) returns the same prediction at locations where $s_m$ is the same. Thus, an additional input feature $q_3$ should be introduced, which must be rotational/reflectional, extended Galilean and scale-invariant, and satisfy the following relation under any $s(y_A^+) = s(y_B^+)$ as

$$q_3(y_A^+) \neq q_3(y_B^+). \tag{11}$$

From the physical standpoint, an inappropriate turbulent timescale is chosen in Eq. (8). There are two assumptions to derive Eq. (8) from Eq. (7): (i) the feature $\nu$ is excluded as an argument, and (ii) the macro timescale of turbulence $k/\varepsilon$ is chosen as the flow timescale from the wall to the fully turbulent region. Lumley[40] showed that Reynolds stresses are uniquely related to the mean-flow field and macroscale of turbulence for a high-Reynolds-number nearly homogeneous flow. This is why conventional turbulence models offer good predictions only in the regions far from the wall. Jiang et al.[3] made a comparative study to demonstrate this issue. In the near-wall region, a correction is needed to account for the dominant viscous effect. According to the expansions of turbulent velocity field near the wall by Hanjalic and Launder,[71] the turbulent timescale (i.e., the ratio of the length scale of the energy-containing eddies to the turbulent velocity scale) approaches a nonzero value. However, the commonly-used timescale $k/\varepsilon$ vanishes at the wall due to $k \to 0$. The timescale is expected to be the Kolmogorov timescale $\sqrt{\nu/\varepsilon}$ because viscous dissipation dominates near the wall. Thus, we define a new timescale as

$$\tau_I \equiv \sqrt{(k/\varepsilon)^2 + c_t^2(\nu/\varepsilon)} = \sqrt{\lambda}(k/\varepsilon), \quad \lambda \equiv 1 + c_t^2/Re_t, \tag{12}$$

where $c_t (> 0)$ is a weighting parameter and $Re_t \equiv k^2/(\nu\varepsilon)$ is turbulent Reynolds number which is widely used in low-Reynolds-number models.[72-75] $\lambda (\geq 1)$ reflects the ratio of the new timescale to conventional timescale. The new timescale $\tau_I$ has good properties: (i) it recovers to standard $k/\varepsilon$ conventional timescale. The new timescale $\tau_I$ has good properties: (i) it recovers to the standard $k/\varepsilon$ at large $Re_t$, and (ii) it never falls below $c_t\sqrt{\nu/\varepsilon}$ ($Re_t \ll 1$). It is worth noting that $\tau_I$ can prevent the singularity at the wall that results from $k/\varepsilon$ vanishing.

Using $s' \equiv \tau_I S$ and $\omega' \equiv \tau_I \Omega$, Eq. (8) can be rewritten as

$$a = g_1' s' + g_2'(s'^2 - \frac{1}{3}\text{tr}(s'^2)\mathbf{I}) + g_3'(s'\omega' - \omega's') + g_4'(\omega'^2 - \frac{1}{3}\text{tr}(\omega'^2)\mathbf{I}) + \cdots \tag{13}$$

with all closure coefficients determined using $f_1 : q' \mapsto g'$, where $q' = \{s_m', \omega_m', \ldots\}$, $s_m' = \sqrt{\lambda} s_m$, and $\omega_m' = \sqrt{\lambda} \omega_m$. Thus, in two-dimensional fully developed channel flows, $a_{12}$ of Eq. (13) eventually depends on both $s_m$ ($s_m = \omega_m$) and $Re_t$ at a given $c_t$. In the following, we examine whether $Re_t$ confronts the nonunique mapping problem. As Fig. 3(a) shows, at any two locations with the same strain, i.e., $s_m(y_A^+) = s_m(y_B^+)$, there always exist two different values of $Re_t$, i.e., $Re_t(y_A^+) \neq Re_t(y_B^+)$. Similarly, at any two locations with $Re_t(y_A^+) = Re_t(y_B^+)$, there always exist $s_m(y_A^+) \neq s_m(y_B^+)$. That is, $s_m(y_A^+) = s_m(y_B^+)$ and $Re_t(y_A^+) = Re_t(y_B^+)$ do not occur at the same time. The selection of $Re_t$ satisfies the condition of Eq. (11) and successfully overcomes the nonunique input–output mappings in conventional models represented by Eq. (8).

The final question is how to determine the value of $c_t$ in order to directly utilize Eq. (13). We note that $\lambda > 1$ implies a correction to the conventional timescale $k/\varepsilon$. Thus, a different value of $c_t$ produces a different range where the correction is activated. In the viscous sublayer ($y^+ < 5$), $k = cy^2$, where $c$ is inversely proportional to the square of the timescale.[74] We set $c = 1/(c_t^2 \nu/\varepsilon)$ in the vicinity of the wall. The transport equation of $k$ reduces to $\nu \partial^2 k/\partial y^2 = \varepsilon$ at the wall.[76] Combining the above relations yields $c_t = \sqrt{2}$. This derivation implies that a correction takes place only in the viscous sublayer. The variation of $\lambda$ at different values of $c_t$ is further investigated based on DNS in Fig. 3(b). Obviously, $c_t = \sqrt{2}(n = 1)$ corresponds to a correction within $y^+ < 10$ (slight departure from $y^+ < 5$). Similarly, $c_t = 4(n = 4)$ and $c_t = 8\sqrt{2}(n = 7)$ correspond to corrections within $y^+ < 50$ and $y^+ < 500$, respectively. However, there is no *prior* knowledge about the exact upper bound below which a correction is applied. Consequently, a strictly reasonable value of $c_t$ may not be available. Thus, Eq. (13) cannot be directly used for turbulence modeling because the input features $\boldsymbol{q}' = \{s_m', \omega_m', ...\}$ and $\boldsymbol{Q}' = \{s', \omega'\}$ are undetermined. To overcome this difficulty, we reform Eq. (13) as

$$\boldsymbol{a} = c_1 \boldsymbol{s} + c_2 (\boldsymbol{s}^2 - \frac{1}{3}\text{tr}(\boldsymbol{s}^2)\mathbf{I}) + c_3 (\boldsymbol{s}\boldsymbol{\omega} - \boldsymbol{\omega}\boldsymbol{s}) + c_4 (\boldsymbol{\omega}^2 - \frac{1}{3}\text{tr}(\boldsymbol{\omega}^2)\mathbf{I}) + \cdots \quad (14)$$

where the closure coefficients $c_1 \equiv \sqrt{\lambda} g_1'$, $c_2 \equiv \lambda g_2'$, $c_3 \equiv \lambda g_3'$ and $c_4 \equiv \lambda g_4'$ can eventually be expressed as functions of $\{s_m, \omega_m, ...\}$ and $Re_t$ if $c_t$ is known. The advantage of Eq. (14) is that the dependence of Eq.(13) on $c_t$ is entirely imputed to the closure coefficients $\{c_i\}_{i=1}^{10}$. The unknown parameter $c_t$ can be further implicitly considered by the ML-based modules when using $\{s_m, \omega_m, ...\}$ and $Re_t$ as input features to directly regress $\{c_i\}_{i=1}^{10}$. The problem is now closed. Notably, although Eq. (14) is directly used for data-driven turbulence modeling, its physical meaning lies in Eq. (13), which interprets why $Re_t$ should be included as an input feature.

### E. Enforcing realizability constraints to outputs

As noted previously, realizability is a physically valuable guiding principle in developing turbulence models which is useful in numerically stable predictions. Efforts have been made to achieve realizable solutions in conventional models,[77-79] however few attempts are made in data-driven turbulence modeling. We directly enforce realizability constraints to the targeted stress anisotropy rather than to closure coefficients themselves. An implementation on closure coefficients may be practicable in conventional models, but considerably difficult in data-driven models because closure coefficients are "implicitly" represented by an ML-based module. On the other hand, for data-driven models directly reconstructing the stress anisotropy, it is also needed to directly implement realizability constraints on the stress anisotropy. Thus, we develop a general procedure.

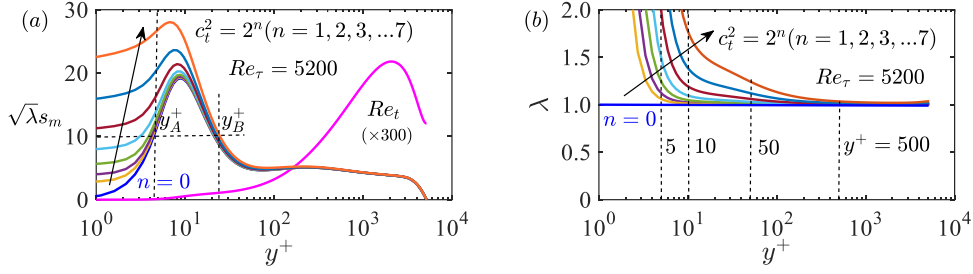

FIG. 3. Two-dimensional fully-developed channel flow at $Re_\tau = 5200$:[70] (a) the shear parameter $s_m$ and turbulent Reynolds number $Re_t$ along the wall-normal direction $y^+$, and (b) the ratio of new turbulent timescale to macro timescale along the wall-normal direction $y^+$. Results at different values of the weighting parameter $c_t$ are shown.

Schumann[80] showed that realizability constraints contain a set of inequalities, however they are unfriendly to provide an effective correction strategy in computational codes. Instead, we select the following inequalities

$$a_{\beta\beta} \geq -\frac{1}{3}, \quad a_{\beta\gamma}^2 \leq (a_{\beta\beta} + \frac{1}{3})(a_{\gamma\gamma} + \frac{1}{3}), \tag{15a,b}$$

$$\lambda_1 \geq \frac{3|\lambda_2| - \lambda_2}{2}, \quad \lambda_1 \leq \frac{1}{3} - \lambda_2, \tag{15c,d}$$

where $\lambda_i (i=1,2,3)$ are eigenvalues of the stress anisotropy $\boldsymbol{a}$ ($\lambda_1 \geq \lambda_2 \geq \lambda_3$). It is noted that no summations are taken for Greek indices. The above conditions are mathematically rigorous and physically important. The first condition Eq. (15a) prevents negative energy. The second condition Eq. (15b) originates from the Schwarz inequality and offers the upper bound for the shear stresses. The remaining conditions Eqs. (15c)-(15d) are supplied to ensure the Reynolds stress positive semidefinite. Thus, Eqs. (15a)-(15d) are sufficient to satisfy the realizability constraints.

Despite selecting constrained conditions, a correction strategy should be further developed when Eqs. (15a)-(15d) are not completely satisfied. Accordingly, we propose a progressive iteration realizability (PIR) scheme as the corrector which is detailed in algorithm 1. The PIR scheme is applied to both training and testing phases, thus achieving a physically consistent and numerically stable data-driven model. In comparison, conditions adopted by Ling et al.[16] are under-constraint. A comparative testing will be conducted in Section IV.A.

**ALGORITHM 1**: Progressive Iteration Realizability (PIR) Scheme.

**Inputs**: Stress anisotropy tensor $\boldsymbol{a}$.

$\boldsymbol{a} \leftarrow (\boldsymbol{a} + \boldsymbol{a}^T)/2$;

**for** $N$ iterations **do**

    **while** $\min\{a_{\beta\beta}\} < -1/3$ **do**

        $a_{\beta\beta} \leftarrow -a_{\beta\beta}[\min\{a_{\beta\beta}\}]^{-1}/3$; // narrow nonzero trace

    **end**

    **while** $a_{\beta\gamma}^2 > (a_{\beta\beta}+1/3)(a_{\gamma\gamma}+1/3)$ **do**

        $a_{\beta\gamma} \leftarrow \sqrt{\max\{(a_{\beta\beta}+1/3)(a_{\gamma\gamma}+1/3),0\}} \cdot \text{sign}\{a_{\beta\gamma}\}$;

    **end**

    Compute eigenvalues of $\boldsymbol{a}$: $\lambda_1 \geq \lambda_2 \geq \lambda_3$;

    **while** $\lambda_1 < (3|\lambda_2|-\lambda_2)/2$ **do**

        $\lambda_1 \leftarrow \lambda_1(3|\lambda_2|-\lambda_2)/2\lambda_1$; // as an amplifier

    **end**

    **while** $\lambda_1 > 1/3 - \lambda_2$ **do**

        $\lambda_1 \leftarrow \lambda_1(1/3-\lambda_2)/\lambda_1$; // as a reducer

    **end**

    Reconstruct $\boldsymbol{a}'$ with updated eigenvalues and initial eigenvectors;

    $a_{\beta\gamma} \leftarrow (a'_{\beta\gamma} + a'_{\gamma\beta})/2$ $(\beta \neq \gamma)$; // only update off-diags to retain zero trace

**end**

**Outputs**: Return well-constrained $\boldsymbol{a}$.

## III. Training: Data Preparation and Learning Strategy

In this phase, we provide a systematic procedure to train a data-driven turbulence model using ML algorithms, especially focusing on how to design a fair and robust learning strategy in order to achieve unbiased and noisy-insensitive predictions.

Fairness in data-driven turbulence modeling is a completely new concept although it has been an important consideration in many fields, such as high-risk decision-making scenarios. As a domain-specific concept, fairness herein refers to the *absence of any prejudice or favoritism in predicted outcomes toward specific flow domain, stress component and coordinate system*. However, the problem of unfairness is essentially the same in data science, which arises from the hidden or neglected biases in selection of training data or objective-function design. For instance, we expect to improve the predictive performance of a model in a certain flow domain, however we may use fewer data in this flow-domain to train this model. It is a simple truth that less data means less contribution to the overall training error and thus lower accuracy. Obviously, fairness is of critical importance especially for data-driven turbulence modeling because obtaining a data-driven model heavily depends on the training strategy based on training data. We present two kinds of fair measures, i.e., a clustering-based data sampling technique and a fair objective-function design.

Another concern is the robustness of a data-driven model. Robustness herein refers to the *insensitivity of a data-driven model to a certain degree of noise or perturbation in input features*. To demonstrate this problem, we conduct a brief review about how it is settled down in conventional turbulence models. Taking the explicit algebraic stress model developed by Gatski and Speziale[81] as an example, the turbulent viscosity can be expressed as follows:

$$\frac{3}{3-2c_\eta^2+c_\xi^2} \approx \frac{3(1+c_\eta^2)}{3+c_\eta^2+6c_\xi^2 c_\eta^2+6c_\xi^2} \approx \frac{1}{1+4c_\eta^2+c_\xi^2}, \tag{16}$$

where $c_\xi$ and $c_\eta$ are strain-dependent coefficients. These three algebraic relations are nearly identical for turbulent flows that are close to equilibrium. However, the right-hand side[82] of Eq. (16) exhibits the best numerical stability because the resulting model responds more insensitively to the small perturbations in the strain. The key lies in noisy-insensitivity of modified turbulent viscosity to strain-dependent coefficients. Jiang et al.[3] summarized the general principle for conventional explicit turbulence models. In comparison, there have not been such considerations in "implicit" data-driven turbulence modeling. Our work is an initial attempt.

### A. Representative dataset for training and testing

Representative datasets of turbulent flows should be carefully selected to investigate the predictive skills of a data-driven turbulence model in capturing different physical phenomena in turbulent flows. The two-dimensional fully-developed channel flow is a general representative of pressure-driven wall-bounded flows. This flow is dominated by the shear stress and has nonlocal effects. Most conventional turbulence models cannot offer approximate predictions in near-wall region. We use this simple parallel shear flow to demonstrate whether additional input feature $Re_t$ can capture the flow characteristics near the wall. Four DNS datasets at various friction Reynolds numbers (based on the friction velocity $u_\tau$ and half-channel height $h$) are used: (i) $Re_\tau = 650$ from Iwamoto et al.,[83] (ii) $Re_\tau = 1000$ from Bernardini et al.,[84] (iii) $Re_\tau = 4200$ from Hoyas and Jimenez,[85] and (iv) $Re_\tau = 5200$ from Lee and Moser.[70] Additionally, three-dimensional fully-developed duct flow is also selected as a canonical example due to its rich characteristics (e.g., secondary vortical structure) and extensive engineering interest. This flow is homogeneous in the streamwise direction, $z$ and $y$ denote the directions parallel to the horizontal and vertical walls. Six DNS datasets at various $Re_\tau$ and aspect ratios (AR) are used:[86,87] (i) AR=1, 3, 5, 7 at $Re_\tau = 180$ and (ii) AR=1, 3 at $Re_\tau = 360$. Both training and testing datasets are from these two flows.

To utilize data from different sources, the raw data should be preprocessed using a unified standard. Thus, we reformat the data and compute the input feature $Re_t$. For channel flows, all raw data are nondimensionalized by $u_\tau$ and the viscous length scale $\nu/u_\tau$ (signified with +); thus, $k^+ = k/u_\tau^2$, $\varepsilon^+ = \nu\varepsilon/u_\tau^4$, and $Re_t \equiv k^2/(\nu\varepsilon) = k^{+2}/\varepsilon^+$ (see Fig. 3). For duct flows, all raw data are normalized by the bulk velocity $u_b$ and $h$ (signified with *); thus, $k^* = k/u_b^2$ and $\varepsilon^* = h\varepsilon/u_b^3$. Accordingly, $Re_t = Re_b k^{*2}/\varepsilon^*$, where $Re_b \equiv u_b h/\nu$ is the bulk Reynolds number. Other input features and the targeted stress anisotropy are computed similarly. Representative input features $Re_t$ and $s_m$ for duct flows are shown in Fig. 4. Note that $s_m = \omega_m$ in channel flows and $s_m \approx \omega_m$ in duct flows, so the redundant feature $\omega_m$ can be omitted.

### B. Clustering-based unbiased sampling technique

It is common knowledge that an ML-based training process could exhibit a preference toward a pattern that is shared by more training data, especially when adopting the commonly-used mean squared error as the objective-function. As a result, an unfair data-driven model is achieved. TBNN-1 in Fig. 2(b) is just the case, where TBNN-1 provides better predictive performance in the region far from the wall because there is more training data in this region. A "pattern" that covers more data accumulates larger error and thus makes bigger impact; in return this pattern is taken high-priority care of by the error backpropagation. Thus, an unbiased strategy is needed to ensure equal sample diversity from training data.

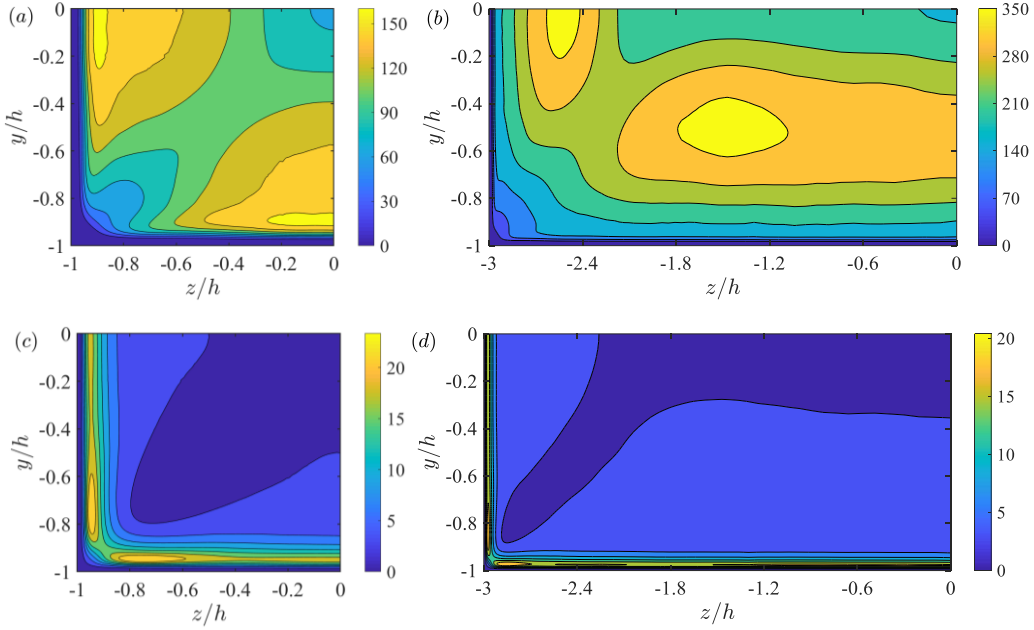

FIG. 4. Representative input features (a)-(b) $Re_t$ and (c)-(d) $s_m$ for duct flows at (a)-(c) $Re_\tau = 180$ with $AR = 1$ and (b)-(d) $Re_\tau = 360$ with $AR = 3$. Only lower left quadrant of the duct is shown due to symmetry.

A k-means clustering algorithm[88] as an unsupervised ML classifier is employed here to assign the data into groups without prior knowledge. Considering that the stress anisotropy is targeted in the objective-function, the stress anisotropy is selected as flow features for clustering. Consequently, the k-means algorithm automatically categorizes the flow whole domain into two different subregions, as shown in Fig. 5, resulting in similar patterns within the same cluster, i.e., nearly homogeneous flows in a region far from the wall and nonhomogeneous flows in a region near the wall. Conventional turbulence models, as demonstrated by Jiang et al.[3], cannot account for near-wall flows. This clustering result, not surprisingly, reflects the effects of wall proximity and can be explained using layered eigenvalues of the stress anisotropy. As shown in Fig. 6, the spatial distribution of each eigenvalue is consistent with the clustering result in Fig. 5. The major role of adopted k-means algorithm is to quantitatively characterize the boundary between two subregions, which cannot be qualitatively extracted from Fig. 6.

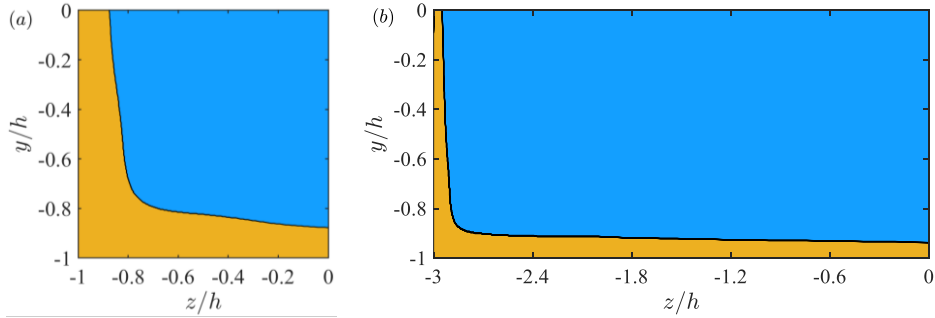

FIG. 5. Clustering results for duct flows at (a) $Re_\tau = 180$ with $AR = 1$ and (b) $Re_\tau = 360$ with $AR = 3$. The whole domain is separated into two subregions: regions far from the wall (blue) and near the wall (yellow).

On the other hand, for two-dimensional channel flows, the clustering boundary approximately locates where the maximum shear stress takes place with a scaling law $Re_\tau^{3/10}$. Chen et al.[89] derived theoretically this place has a scaling law of $Re_\tau^{1/3}$. From the above discussion, it can be seen that a clustering-based data sampling technique for data-driven turbulence modeling is not only a mathematical operation but also has a physical meaning. Accordingly, an oversampling technique based on this clustering technique is developed to rebalance equal sample diversity between two clustered subregions, and thus guarantee that different flow patterns are treated fairly in the training process. It is worth noting that the clustering-unbiased sampling technique herein will be very useful to develop a universal data-driven model across multiple classes of flows in the near future.

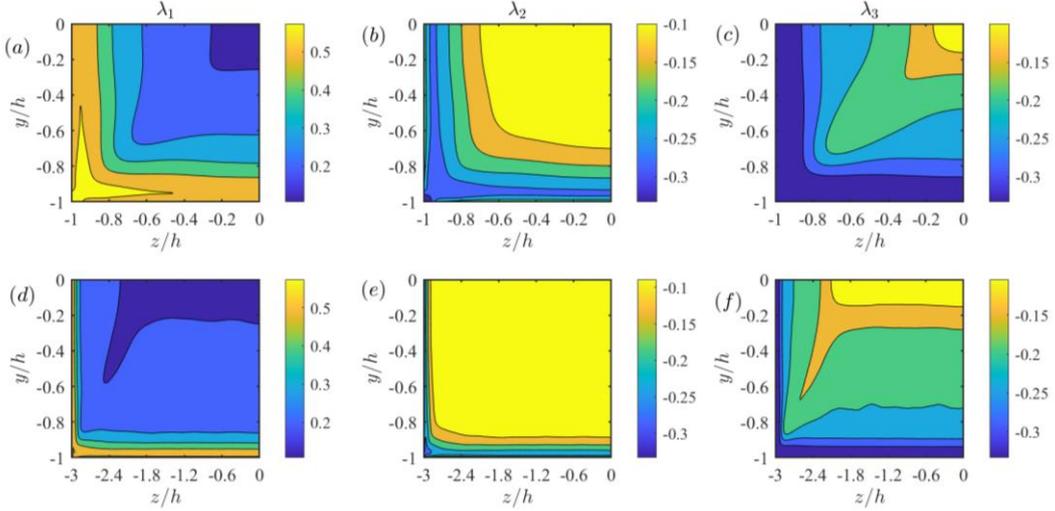

FIG. 6. Eigenvalues ($\lambda_1$, $\lambda_2$, $\lambda_3$) of the stress anisotropy for duct flows at (a)-(c) $Re_\tau = 180$ with $AR = 1$ and (d)-(f) $Re_\tau = 360$ with $AR = 3$. The first to third columns corresponds to $\lambda_1$, $\lambda_2$, and $\lambda_3$, respectively. Only lower left quadrant of the duct is shown due to symmetry.

## C. Neural network design and training experiments

Under proposed UIML shown in Fig. 1, we use $\boldsymbol{q} = \{Re_t, s_m, \ldots\}$ and $\mathcal{Q} = \{\boldsymbol{s}, \boldsymbol{\omega}\}$ as input features and constrain the learned integrity bases to conventional tensor bases of Pope.[41] This simplification aims to create a controlled environment in which main influence factors regarding UIML are well behaved and thus the dynamical properties of UIML can be explored in isolation in the near future. Moreover, data-driven inverse modeling under a set of constrained integrity bases is a proof-of-concept of UIML as a transfer learning framework. In theory, Eq. (8) is valid only under the assumption of homogeneous strain rates. Here we show how to transfer it to nonhomogeneous flows (e.g., near-wall flows in a channel and secondary flows within a duct) with the aid of newly-proposed $Re_t$. With these in mind, we develop our data-driven model.

The main focus in the training phase is thus on the multicomponent regression problem defined as $f: \{\boldsymbol{q}, \mathcal{Q}\} \mapsto \boldsymbol{a}$ under a set of targeted needs, i.e., requirements of a fair prediction for different flow domains and stress components, noisy-insensitivity of learned model to input features, spatial smoothness of predicted field, physical realizability of predicted components, and extrapolation capability. These needs are significantly important for physical problems. An ML-based procedure to settle down this problem herein integrates three critical steps: (i) setup a proper neural network architecture to provide a family of alternative functions with a set of trainable network parameters; (ii) design a reasonable objective-function to fully consider targeted needs; and (iii) select a good optimization algorithm to determine the trainable parameters.

A specialized physics-informed residual network (PiResNet) under UIML is developed to represent the mapping $f:\{q, \mathcal{Q}\} \mapsto a$ on the training dataset $\mathcal{D}$, as shown in Fig. 7. PiResNet is based on a residual network architecture with fully-connected layers. This network is an acyclic cascade consisting of an input layer, several hidden layers, and an output layer. A skip connection is applied across several hidden layers to provide identity mapping, thus forming a residual block in which the overall operation can be expressed as $F(x_i) + x_i \mapsto x_{i+m}$ and $F(\cdot)$ is the overall operation of $m$ fully-connected layers contained in this residual block. In each fully-connected layer, $\sigma(w_i x_i + b_i) \mapsto x_{i+1}$, where $w_i$ and $b_i$ are trainable network parameters and $\sigma$ is the nonlinear activation function. He et al.[46,47] has demonstrated, in image recognition, that the identity mapping achieves great empirical success to settles down two technical problems frequently encountered in deep learning, i.e., gradient vanishing and performance degradation. Considering the fact that there is little use of other networks other than plain neural network in turbulence closure modeling, the first use of this network enriches our understanding of neural networks. Recalling PiResNet, we cast the form-constraint inverse modeling as an optimization problem by minimizing the following objective-function

$$\mathcal{J}(\theta; \lambda_w, \lambda_c; \mathcal{D}) \equiv \mathcal{L}_a + \lambda_w \mathcal{L}_w + \lambda_c \mathcal{L}_c, \tag{17a}$$

$$\mathcal{L}_a \in \left\{ \mathcal{L}_a^1 \equiv \frac{1}{9N} \sum \| a - \tilde{a} \|_F^2, \; \mathcal{L}_a^2 \equiv \frac{1}{9N} \sum \| \gamma \odot (a - \tilde{a}) \|_F^2, \; \mathcal{L}_a^3 \equiv \frac{1}{9N} \sum \| a \tilde{a}^{-1} - \mathbf{I} \|_F^2 \right\}, \tag{17b}$$

$$\mathcal{L}_w \equiv \| w \|_2^2, \quad \mathcal{L}_c \equiv \frac{1}{4N} \sum \| c \|_2^2, \tag{17c}$$

where $\theta = [w, b]$ are trainable parameters, $N$ is the number of samples in $\mathcal{D}$, $\lambda_w$ and $\lambda_c$ are regularization factors, and $\gamma$ is a fair-factor matrix. $\mathcal{L}_a$ represents supervised data-driven aspects, for which we provide three alternatives in Eq. (17b) to achieve an unbiased prediction for the stress anisotropy components. $\mathcal{L}_w$ represents the complexity of the network; thus, $\lambda_w \neq 0$ to prevent overfitting and obtain noise-insensitive closure coefficients. The regularization constraint on $\mathcal{L}_c$ is intended to ease ill-posed problems and reduce the noise-sensitivity of the learned model to the structural bases. Both $\lambda_w \neq 0$ and $\lambda_c \neq 0$, as well as the PIR scheme constrained on $a$, are unsupervised physics-informed aspects embedded in the training phase. Finally, the Adam algorithm[90] is adopted to compute the cost function gradients, and the backpropagation algorithm[58] is used to update the network parameters. The training procedure is summarized in Algorithm 2, and serves as a fair and robust training strategy to achieve optimal parameters $\theta_\pi$; thus, $a = f(q, \mathcal{Q}; \theta_\pi)$.

The numeric range of input features and outputs should be limited in order to ease the updating of trainable network parameters. A logarithmic function $\log(1 + x)$ is selected to rescale all input features, which makes $Re_t$ work better. In comparison, the learned model responses more sensitively to noise in input features when adopting commonly-used normalization with global means and variances, which should be avoided. We note that the benefits of logarithm functions may go far beyond feature scaling. Weatheritt and Sandberg[13] found that the logarithm has been always successfully evolved in every best regression solution when using evolutionary algorithm. At the same time, integrity bases are also rescaled in order to adjust numeric range of learned closure coefficients, as shown in Fig. 7. Previous work by Jiang et al.[3] demonstrated that closure coefficients do have a separation of order magnitude. Another concern is the selection of activation functions, which is based on following considerations: (i) smooth functions (infinitely differentiable) to account for the smoothness requirement of predicted stress anisotropy and (ii) sensitivity to negative values to prevent dead neurons. Thus, we choose Gaussian error linear unit (GELU)[91] as the activator, which does outperform rectified linear unit (ReLU)[92] and exponential linear unit (ELU)[93] in our testing. GELU achieves smaller network weights and thus is more insensitive to noise in input features. Good properties of a gating mechanism by GELU are detailly investigated.[94] We note that PiResNet training with GELU has a little higher computational cost, almost 8.4 GPU hours (GeForce GRT 2080). Besides, a comparative testing of plain neural network (16 hidden layers) and residual neural network (10 hidden layers) indicates that the latter achieves better and more robust predictions. Thus, residual neural network is selected as the resulting network. PiResNet training is performed with the open-source software library TensorFlow.

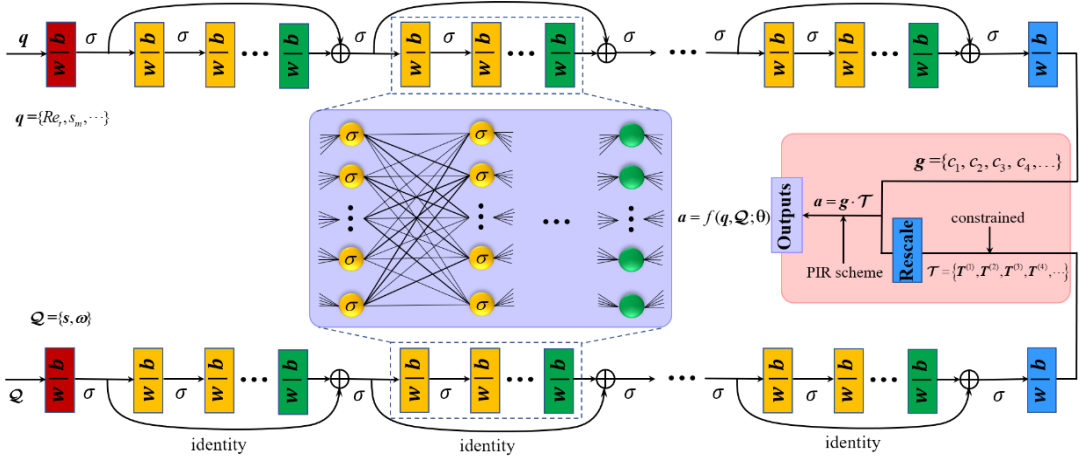

FIG. 7. A schematic of invariant and realizable PiResNet based on a residual network architecture with fully-connected layers. All operations in the neural network are element-wise. A fair and robust training strategy for PiResNet is detailed in algorithm 2.

**ALGORITHM 2:** PiResNet Training Scheme. All operations are element-wise.

MAIN
**Inputs**: Training dataset $\mathcal{D}$, learning rate $\eta$, and weighting factors $(\lambda_w, \lambda_c)$.
**Initialization**: Set $t = 0$;
Randomly initialize $\theta_t = [w, b]$.
**Iteration**: $\mathcal{J}(\theta_t) \leftarrow \text{FORWARD}(q, \mathcal{Q}; \theta_t)$;
while *not convergent* do
 (*cross-validation as a stopping criteria*)
 $t \leftarrow t + 1$;
 $\theta_t \leftarrow \text{BACKWARD}(q, \mathcal{Q}; \theta_{t-1})$;
 $\mathcal{J}(\theta_t) \leftarrow \text{FORWARD}(q, \mathcal{Q}; \theta_t)$;
end
**Outputs**: Return resulting $\theta$ as the optimal $\theta_\pi$.

FORWARD$(q, \mathcal{Q}; \theta)$
**Require**: $q, \mathcal{Q}, \tilde{a}; \lambda_w, \lambda_c; \theta$.
 $a \leftarrow f(q, \mathcal{Q}; \theta)$ as Fig. 7;
 Compute $\mathcal{J}(\theta)$ by Eq. (17).
end

BACKWARD$(q, \mathcal{Q}; \theta)$
**Require**: $q, \mathcal{Q}, \tilde{a}; \lambda_w, \lambda_c, \eta; \theta$.
 Compute $\widetilde{\nabla \mathcal{J}(\theta)}$ using Adam;
 (*gradient with bias-corrected momentum*)
 $\theta \leftarrow \theta - \eta \widetilde{\nabla \mathcal{J}(\theta)}$. // *update rule*
end

The effect of the training dataset size on the performance of PiResNet is investigated with the cross-validation technique, as shown in Fig. 8(a). Three flow cases are used for training: a channel flow at $Re_\tau = 1000$ and two duct flows with AR=1 at $Re_\tau = 180$ and AR=3 at $Re_\tau = 360$; and the remaining cases are used for testing, as shown in Fig. 8(b). It is noted that PiResNet trained only with a duct flow with AR=1 at $Re_\tau = 180$ still achieves a good extrapolation to a duct flow with AR=3 at $Re_\tau = 360$, but slightly underpredicts shear stresses for a duct flow with AR=7 at $Re_\tau = 180$. In comparison, the random forest model by Wang et al.[28] trained with three duct cases at $Re_b$ =2200, 2600 and 2900 (respectively corresponding to $Re_\tau$ =150, 175 and almost 200) extrapolates to a duct flow at $Re_b$=3500 ($Re_\tau$=225); the neural network model by Ling et al.[16] trained with six classes of different flow configurations (including a duct flow at $Re_b$=3500) transfers to a duct flow at $Re_b$=2000 ($Re_\tau$=120). PiResNet does benefit from embedded prior domain-knowledge. Recalling data-driven aspects, the additional duct data helps capture the effect of large aspect ratio better while the additional channel flow helps further improve predictions at duct centerplane and capture the effect of large Reynolds number better.

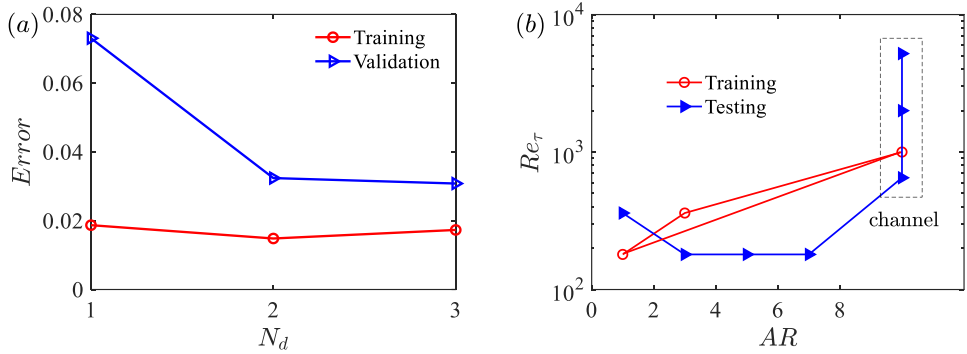

FIG. 8. (a) Training and validation errors (root mean squared error) as functions of the training dataset size ($N_d$) and (b) The flow configurations adopted for PiResNet training and testing.

We use both original data and clustering-based rebalanced data to train PiResNet (denoted as PiResNet-1 and PiResNet-2) and compare the fair prediction ability for two clustering subregions. The discrepancy of prediction errors between these two subregions are shown in Table I, where a smaller value means a fairer prediction. PiResNet-1 trained with imbalance data tends to have an unfair prediction and this unfair trend increases with increasing data imbalance. The ratio of sample numbers of two subregions in original data, i.e., $r_s$, is also shown in Table I to characterize the degree of data imbalance. In comparison, PiResNet-2 with rebalanced data does gain fairer predictions for different flow domains: a higher than 40% improvement in training dataset and a almost 25% improvement in testing dataset, especially a 69% rise for the case with AR=1 at $Re_\tau = 360$. Thus, the clustering-based sampling technique is effective. It is noted that an absolutely fair PiResNet requires a more accurate clustering result to guarantee absolutely equal diversity of different patterns. Despite not absolutely fairness, the clustering-based sampling technique greatly mitigates the bias of a learned model toward a certain flow pattern which will play an important role in develop a universal data-driven model across multiple classes of flows in the near future.

Similarly, we further examine whether PiResNet achieves fair predictions for stress anisotropy components. Eq. (17b) provides three different objective functions for the first term of Eq. (17a), i.e., $\mathcal{L}_a^1$, $\mathcal{L}_a^2$, and $\mathcal{L}_a^3$. In particular, the fair factor in $\mathcal{L}_a^2$ is properly selected by trial-and-error method as $\gamma_{ij} = \sqrt{N/\sum|\tilde{a}_{ij}|}$. The comparative result is shown in Table II. $\mathcal{L}_a^1$ shows biased predictions for $a_{11}$ and $a_{12}$, while $\mathcal{L}_a^2$ and $\mathcal{L}_a^3$ make improvements toward a fair prediction of all stress anisotropy components. In particular, the prediction of $a_{23}$ is significantly improved. However, $\mathcal{L}_a^2$ is not frame-invariant and depends on the matrix $\gamma$ (not a tensor). For the same flow even as the training case, PiResNet with $\mathcal{L}_a^2$ provides different predictive accuracy when only different meshes are used. Thus, we choose $\mathcal{L}_a^3$ in order to form a scale- and frame-invariant objective-function. Consequently, the hyper-parameters can be valid across training datasets of dynamically similar flows only with different scales or different coordinate systems.

TABLE I. The discrepancy of root mean squared errors between two clustering subregions. PiResNet-1 and PiResNet-2 refer to PiResNet trained with original data and clustering-based rebalanced data. $r_s$ denotes the ratio of sample number between two subregions in original data. The comparison of both models is conducted using the same training flows at almost the same level of the overall training error.

|  | AR=1 $Re_\tau=180$ | AR=3 $Re_\tau=360$ | AR=1 $Re_\tau=360$ | AR=3 $Re_\tau=180$ | AR=5 $Re_\tau=180$ | AR=7 $Re_\tau=180$ |
|---|---|---|---|---|---|---|
| $r_s$ | 3.07 | 3.17 | 2.68 | 3.28 | 3.43 | 3.61 |
| PiResNet-1 | $5.97\times10^{-3}$ | $8.85\times10^{-3}$ | $3.33\times10^{-3}$ | $1.96\times10^{-2}$ | $2.01\times10^{-2}$ | $2.05\times10^{-2}$ |
| PiResNet-2 | $3.22\times10^{-3}$ | $5.17\times10^{-3}$ | $1.02\times10^{-3}$ | $1.45\times10^{-2}$ | $1.53\times10^{-2}$ | $1.57\times10^{-2}$ |

TABLE II. The regression coefficients for three different $\mathcal{L}_a$ defined in Eq. (18b) corresponding to $\mathcal{L}_a^1$, $\mathcal{L}_a^2$, and $\mathcal{L}_a^3$, respectively. Results both for all training and testing flows are given.

|  |  | $a$ | $a_{11}$ | $a_{22}$ | $a_{33}$ | $a_{12}$ | $a_{13}$ | $a_{23}$ |
|---|---|---|---|---|---|---|---|---|
| Training | $\mathcal{L}_a^1$ | 0.955 | 0.972 | 0.959 | 0.967 | 0.945 | 0.924 | 0.689 |
|  | $\mathcal{L}_a^2$ | 0.958 | 0.969 | 0.968 | 0.971 | 0.946 | 0.942 | 0.883 |
|  | $\mathcal{L}_a^3$ | 0.956 | 0.969 | 0.968 | 0.969 | 0.944 | 0.945 | 0.886 |
| Testing | $\mathcal{L}_a^1$ | 0.928 | 0.951 | 0.925 | 0.931 | 0.922 | 0.908 | 0.590 |
|  | $\mathcal{L}_a^2$ | 0.933 | 0.939 | 0.938 | 0.940 | 0.923 | 0.920 | 0.789 |
|  | $\mathcal{L}_a^3$ | 0.932 | 0.938 | 0.937 | 0.938 | 0.921 | 0.925 | 0.794 |

Furthermore, we search for optimal regularization factors $\lambda_w$ and $\lambda_c$ from $10^{-3} \sim 10^{-7}$ and eventually select $\lambda_w = 5\times10^{-7}$ and $\lambda_c = 4\times10^{-6}$. In this setting, PiResNet has smaller values both for the overall network parameters and closure coefficients, as shown in Fig. 9. The regularization constraint only on closure coefficients ($\lambda_w = 0$, $\lambda_c \neq 0$) can hardly limit the network parameters (see Fig. 9(b)) and the regularization constraint only on network parameters ($\lambda_c = 0$, $\lambda_w \neq 0$) make limited impact in closure coefficients (see Fig. 9(c)). Thus, regularization constraints both on network parameters and closure coefficients are necessary. It is worth noting that the regularization of the closure coefficients is critical in two aspects. On the one hand, the ill-posed problem that the objective-function gradients w.r.t. the closure coefficients are insensitive may occur when the strain is zero somewhere in the flow. Thus, we introduce additional information to regularize the solution by penalizing the total magnitude of the closure coefficients. To some extent, it also increases smoothness. On the other hand, the closure coefficients can be regarded as an amplifier to the uncertainty of integrity bases. Smaller closure coefficients reduce the noisy-sensitivity of the learned model to integrity bases. Similarly, smaller network parameters mean lower noisy-sensitivity of strain-dependent closure coefficients. PiResNet with $\lambda_w = 5 \times 10^{-7}$ and $\lambda_c = 4 \times 10^{-6}$ eventually achieves a mean value of 0.025 for each network parameter averaged on all neurons and 0.046 for each closure coefficient averaged on the spatial volume. Section IV.A shows their quantitative effects. Finally, the whole flowchart to search optimal PiResNet in this work can be summarized in Fig. 10. The resulting network has 5 residual blocks, and each block contains 2 fully-connected hidden layers with 128 neurons in each layer.

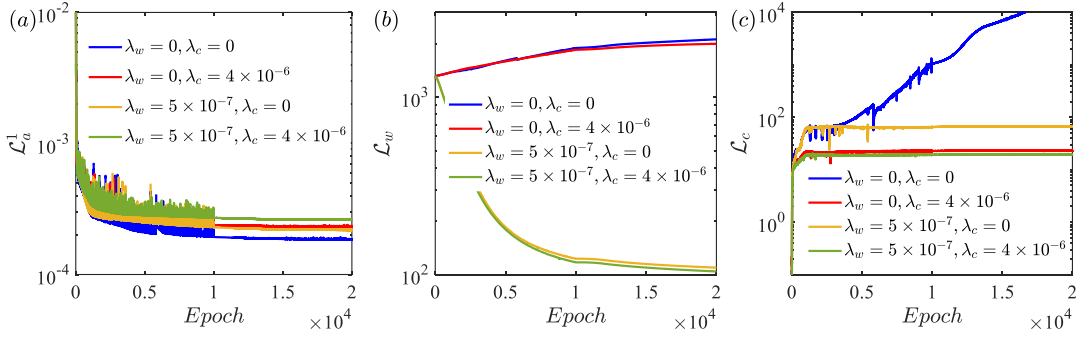

FIG. 9. The learning dynamics of PiResNet with three sets of hyperparameters ($\lambda_w$, $\lambda_c$): (a) $\mathcal{L}_a^1$ representing the quality of data-driven training; (b) $\mathcal{L}_w$ representing the complexity of network; and (c) $\mathcal{L}_c$ representing the magnitude of closure coefficients. All training experiments are performed with $\mathcal{L}_a^3$ and evaluated with $\mathcal{L}_a^1$. Note that $\mathcal{L}_c$ is computed for rescaled closure coefficients due to rescaling integrity bases as shown in Fig. 7.

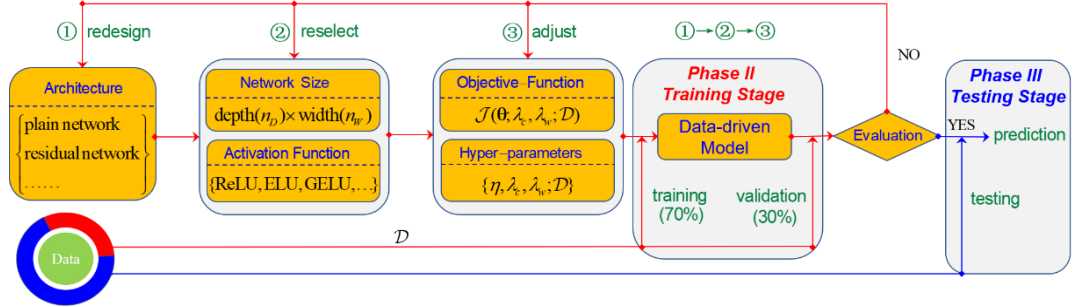

FIG. 10. A flowchart to achieve the optimal PiResNet in this work.

## IV. Testing: Numerical Results and Validation

In this phase, as outlined in Fig. 1, a systematic evaluation of PiResNet-predicted outcomes from different aspects (e.g., extrapolative accuracy, realizability and robustness) is motivated to present evidence driving the trust toward PiResNet. Testing is performed on two classes of different flow configurations: (i) two-dimensional parallel-shear flows in a channel at various Reynolds numbers and (ii) three-dimensional secondary flows within a duct with various aspect ratios. It is well-recognized that conventional closure models lack accuracy both for near-wall flow of a channel (featured with nonhomogeneous effects due to wall proximity) and in-plane flow of a duct (featured with secondary motions induced by normal-stress imbalance). We also note, in these two flows under testing, the roles that Reynolds stress plays in the mean-flow field are different: shear stress dominates the flow in a channel[89] while normal stress controls the secondary flow in a duct.[95] Hence, the utility of PiResNet is demonstrated by applying it to these flows to capture the aforementioned flow characteristics and their sensitivities to Reynolds number and aspect ratio.

To illustrate the superiority of PiResNet, we also compare it to the same-level model based on gene expression programming by Weatheritt and Sandberg[13] (denoted as GEP). It is less fair to compare modern data-driven models to conventional algebraic models, e.g., that is the case in Ref.[16]; instead fair to the state-of-the-art model, at least less outdated models, which is also stressed recently by Musgrave et al.[96] Besides, DNN (as the kernel of PiResNet) and GEP are two representatives of modern ML algorithms: the former is parametric regression based on connectionism while the latter is symbolical regression. These are why we take GEP as the baseline model. This is the first attempt, to our knowledge, to compare different ML-based turbulence models.

### A. Extrapolative predictions of stress anisotropy

The section is centered around whether the well-trained PiResNet can successfully provide satisfactory predictions on unseen flows, i.e., its extrapolative capability. PiResNet-predicted stress anisotropy for channel flows at $Re_\tau$=650, 2000 and 5200 are shown in Fig. 11 and compared with GEP and reference DNS. It can be seen that PiResNet is consistent with DNS in all stress anisotropy components. In comparison, GEP provides satisfactory predictions only for the shear stress and only in the region far from the wall, which is a slight advantage over conventional nonlinear models in this flow. Moreover, GEP, not surprisingly, returns the same stress for near-wall region as the region far from the wall due to inappropriate turbulent timescale. Introducing $Re_t$, which parameterizes the newly-proposed turbulent timescale, does enable PiResNet to extend the resolution scope to the whole flow domain. Thus, the concept of Eq. (12) is effective.

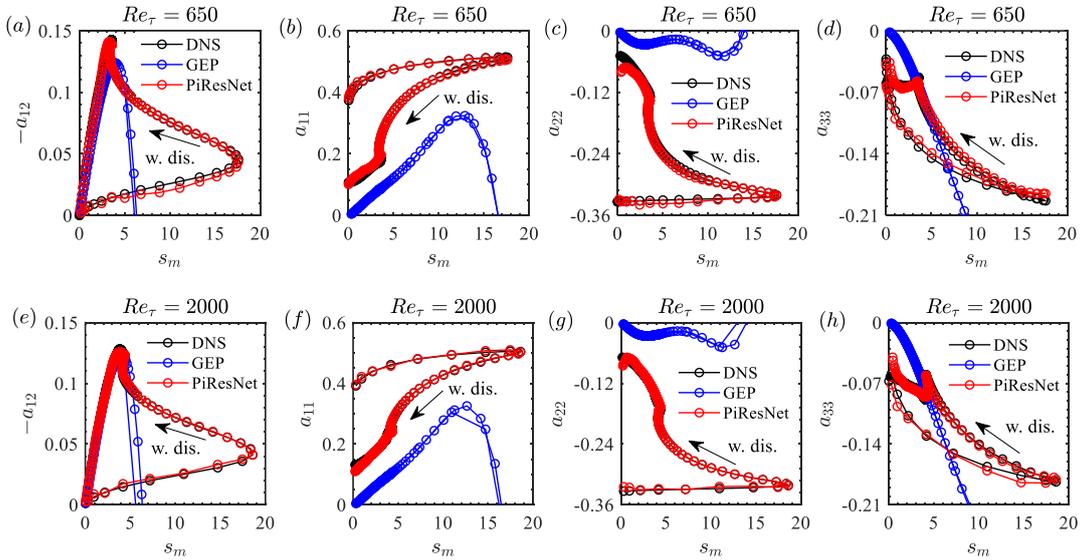

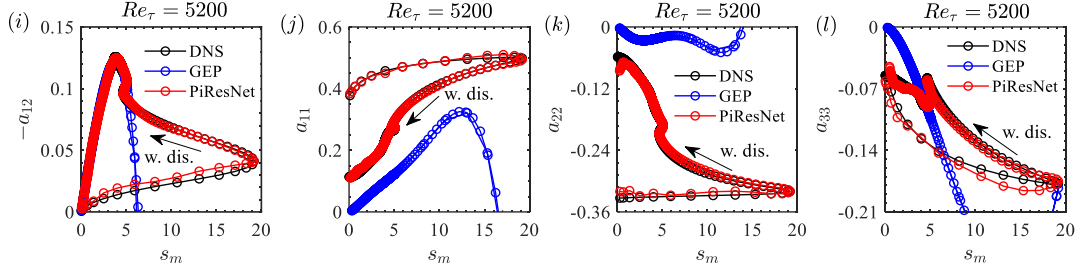

FIG. 11. The variation of stress anisotropy components ($a_{ij}$) with nondimensional shear parameter ($s_m$) predicted by GEP, PiResNet, and DNS for channel flows at (a)~(d) $Re_\tau$=650, (e)~(h) $Re_\tau$=2000, and (i)~(l) $Re_\tau$=5200. The first to forth columns correspond respectively to $a_{12}$, $a_{11}$, $a_{22}$, and $a_{33}$. The direction with increasing wall distance (w. dis.) is also marked. Note that some results of GEP are out of range and not shown.

PiResNet-predicted stress anisotropy for duct flows at $Re_\tau$=360 with AR=1 is shown in Fig. 12 and $Re_\tau$=180 with AR=7 in Fig. 13. For brevity, results for duct flows with smaller aspect ratios (i.e., AR=3 and 5) at $Re_\tau$=180 are omitted. Due to symmetry, only lower left quadrant of the duct ( $-\text{AR} \leq z/h \leq 0, -1 \leq y/h \leq 0$ ) is shown. PiResNet easily provides improvements on predictions of all stress anisotropy components while GEP provides relatively accurate predictions only for shear stresses $b_{12}$ and $b_{13}$. An unfair prediction of normal stresses by GEP lies in its biased training objective-function. It is common knowledge that in-plane normal stress imbalance is of critical importance to capture secondary motions in duct flows. Besides, GEP lacks accuracy in the near-wall region. In comparison, PiResNet correctly captures the secondary mechanism and has a good predictive ability for near-wall flows. PiResNet successfully transfers to flows with different Reynolds numbers and aspect ratios, indicating that PiResNet does learn the underlying similar dynamical mechanism. It is worth noting that a fair training strategy enables PiResNet to improve predictions of $a_{23}$.

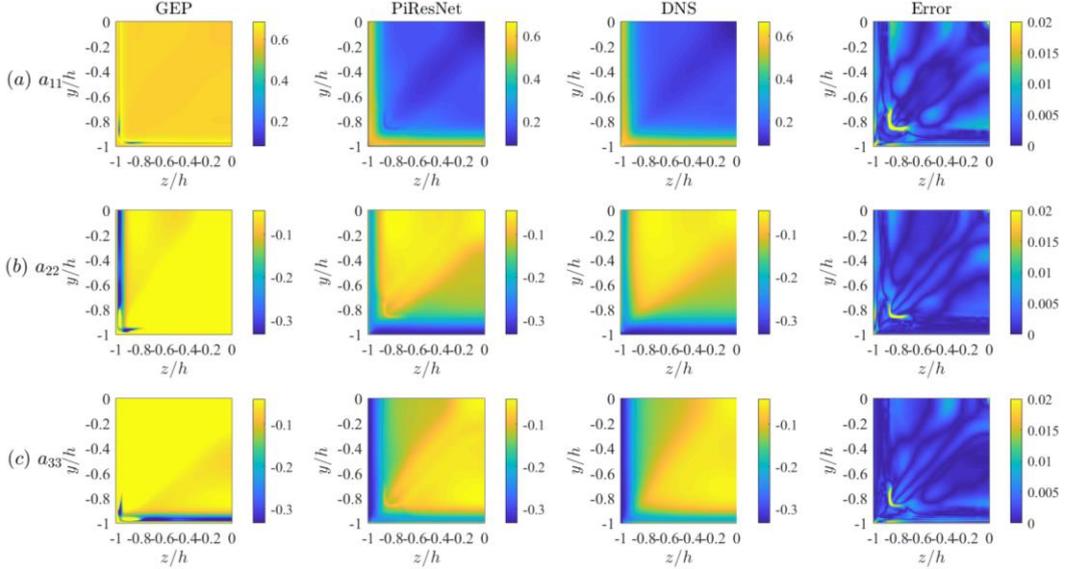

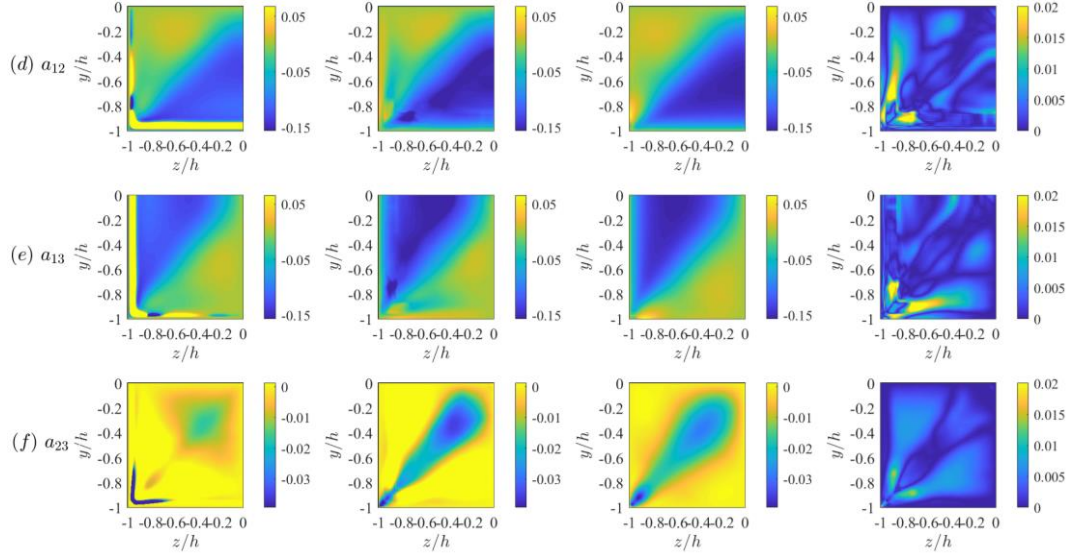

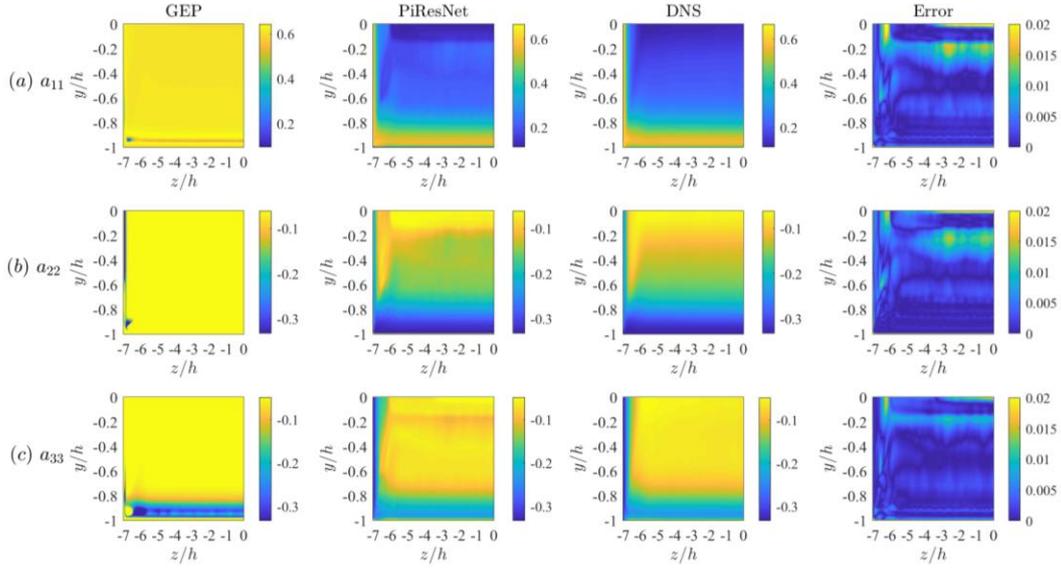

FIG. 12. Stress anisotropy components on the duct cross-plane predicted by GEP, PiResNet, and DNS for a duct flow at $Re_\tau=360$ with AR=1: (a) $a_{11}$, (b) $a_{22}$, (c) $a_{33}$, (d) $a_{12}$, (e) $a_{13}$, and (f) $a_{23}$. The first to forth columns correspond to results of GEP, PiResNet, DNS and the absolute error between PiResNet and DNS. Note that this case is an extrapolation both on Reynolds number and aspect ratio. For simplicity, only lower left quadrant of the duct ($-1 \leq z/h, y/h \leq 0$) is shown. The coordinate origin is located at the duct center.

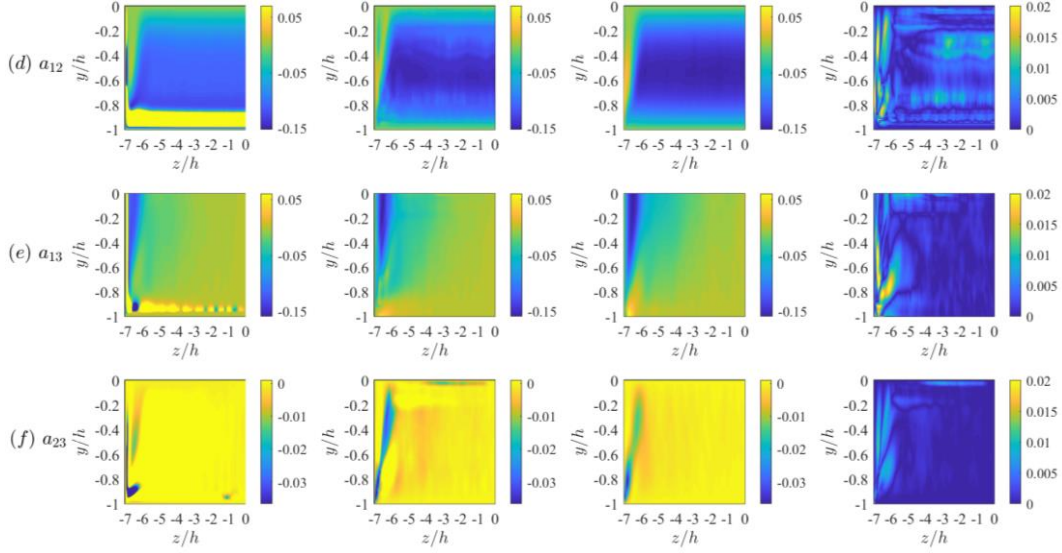

FIG. 13. Stress anisotropy components on the duct cross-plane predicted by GEP, PiResNet, and DNS for duct flow at $Re_\tau=180$ with AR=7: (a) $a_{11}$, (b) $a_{22}$, (c) $a_{33}$, (d) $a_{12}$, (e) $a_{13}$, and (f) $a_{23}$. The first to forth columns correspond to results of GEP, PiResNet, DNS and the absolute error between PiResNet and DNS. Note that this case is a large extrapolation on aspect ratio. For simplicity, only lower left quadrant of the duct ($-7 \leq z/h \leq 0, -1 \leq y/h \leq 0$) is shown. The coordinate origin is located at the duct center.

## B. Realizability capability

We further examine in the barycentric map (BMap)[97] whether PIR corrector really works and whether a realizable solution is achieved by PiResNet. BMap as the non-distorted representation of anisotropy, linearly relates any anisotropy state to three limiting states

$$\boldsymbol{\pi} = \xi_{1c}\boldsymbol{\pi}_{1c} + \xi_{2c}\boldsymbol{\pi}_{2c} + \xi_{3c}\boldsymbol{\pi}_{3c}, \qquad (18)$$

where $\boldsymbol{\pi}_{1c}$, $\boldsymbol{\pi}_{2c}$, and $\boldsymbol{\pi}_{3c}$ denote the three vertices of BMap in representing the one component state (1C), two-component axisymmetric state (2C-axis) and three-component isotropic state (3C-iso). $\xi_{1c}$, $\xi_{2c}$, and $\xi_{3c}$ are weighting factors of these three states, which are determined by the eigenvalues of the stress anisotropy:

$$\xi_{1c} = \lambda_1 - \lambda_2,\ \xi_{2c} = 2(\lambda_2 - \lambda_3),\ \xi_{3c} = 3\lambda_3 + 1. \qquad (19)$$

Thus, comparative results for channel flows at $Re_\tau$=650, 2000 and 5200 are shown in Fig. 14. PiResNet achieves good agreement with DNS and is successfully constrained within the well-defined BMap; however, GEP mismatches with DNS especially in the region far from the wall. GEP shows more obvious bias toward 3C-iso state than PiResNet, thus indicating a poor characterization of anisotropy. Further results for duct flows at $Re_\tau$=360 with AR=1 and $Re_\tau$=180 with AR=7 are shown in Fig. 15 and Fig. 16. It can be seen that PiResNet always provides realizable solutions and agrees well with DNS. In comparison, realizable anisotropy is obtained by GEP only in the region far from the wall and, in particular, GEP returns to a 3C-iso state as the wall is approached. As demonstrated by Weatheritt and Sandberg[13], the qualitative incorrectness is due to inappropriate use of timescale. With Eq. (12), our new timescale $\tau_I$ never tends to zero as the wall is approached.

Without the PIR corrector, our proposed PiResNet provides realizable solutions in most flow region other than the near-wall region. To demonstrate the effectivity of proposed PIR corrector and its advantage over that of Ling et al.,[16] both realizability correctors are applied to GEP-predicted stress anisotropy. As a result, our proposed PIR corrector easily returns realizable solutions while that of Ling et al.[16] is still under-constraint, thus indicating that the PIR corrector is effective and can be extended to other turbulence models for realizable solutions.

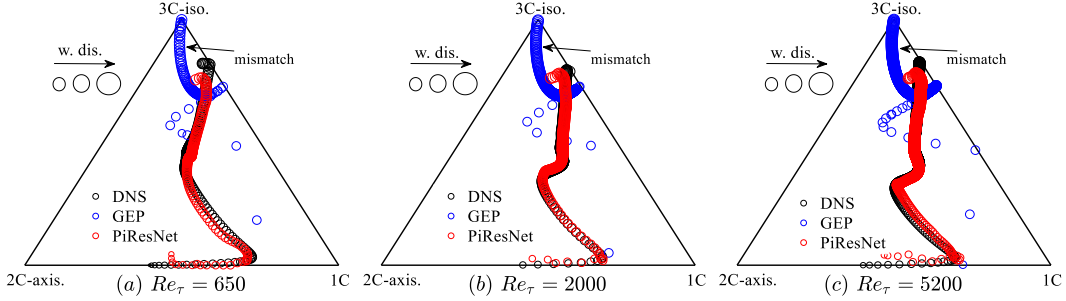

FIG. 14. Scatter plot of the stress anisotropy in the barycentric map for channels flow at (a) $Re_\tau$=650, (b) $Re_\tau$=2000, and (c) $Re_\tau$=5200, scaled with the corresponding wall distance (w. dis.). Results of GEP and PiResNet are compared with DNS. Note that some results of GEP are out of range and not shown.

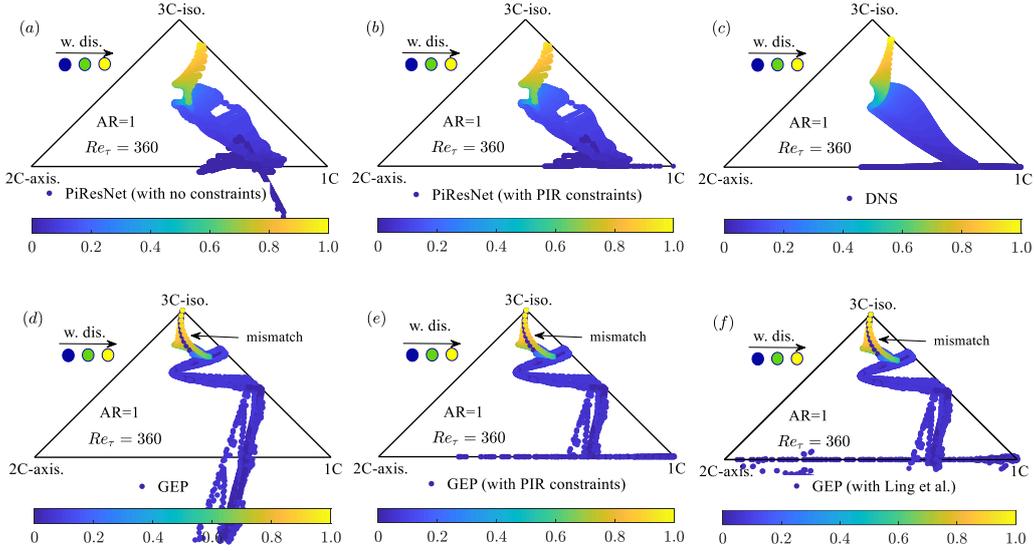

FIG. 15. Scatter plot of the stress anisotropy in the barycentric map for a duct flow at $Re_\tau$=360 with AR=1, colored by the corresponding wall distance (w. dis.). Results of GEP and PiResNet are compared with DNS.

## C. Robustness performance

The robustness of PiResNet is investigated in terms of its noise-sensitivity. Additional Gaussian noise is added to the input features in the form

$$\varphi' = \varphi(1+\delta\ell), \quad \ell \sim \mathbb{N}(0,1), \tag{20}$$

where $\varphi'$ and $\varphi$ denote the input features with and without noise, $\delta$ is the noise level and $\ell$ is a set of random numbers that satisfies a Gaussian distribution. The relative error is defined as

$$\varepsilon_a = \langle \|\boldsymbol{a}-\hat{\boldsymbol{a}}\|_F / \|\hat{\boldsymbol{a}}\|_F \rangle, \tag{21}$$

where $\langle \cdot \rangle$ denotes averaging over the spatial flow domain. Figure 17 shows the variation of relative prediction error of PiResNet with various noise levels in input features. Three activators with the same regularization operation are compared. It can be seen that the prediction error is almost unchanged until a noise level of 30% and PiResNet with GELU is more insensitive to perturbation in input features. Thus, PiResNet gains good noise immunity, which is of critical importance to numerical simulations. Also, it is very useful to improve experimental Reynolds stress when combining PiResNet and noisy experiment data.

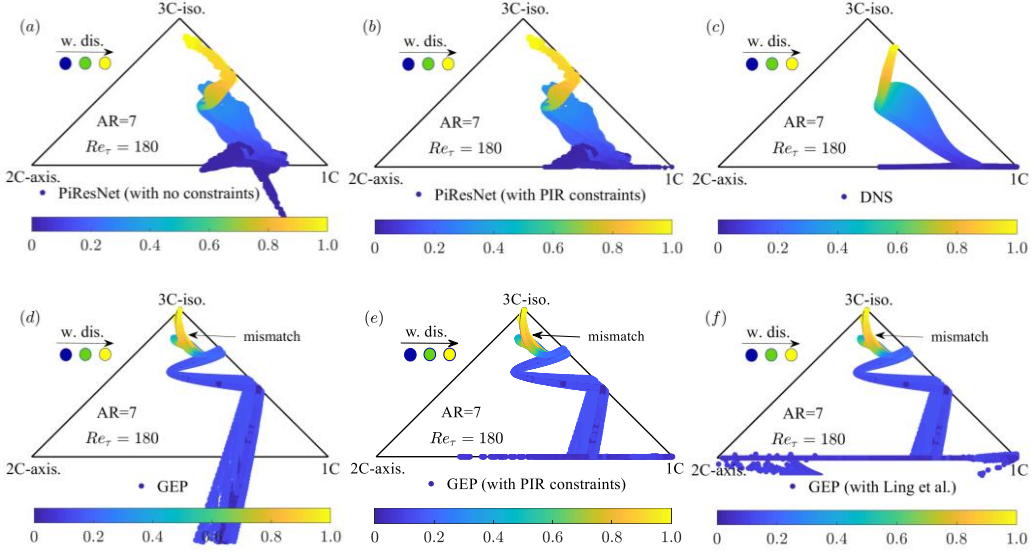

FIG. 16. Scatter plot of the stress anisotropy in the barycentric map for a duct flow at $Re_\tau$=180 with AR=7, colored by the corresponding wall distance (w. dis.). Results of GEP and PiResNet are compared with DNS.

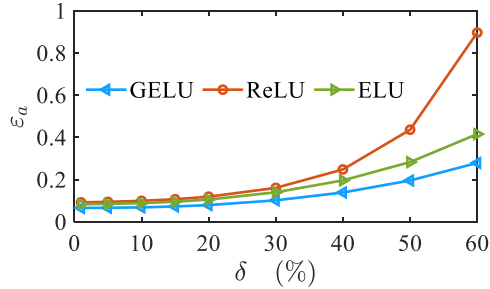

FIG. 17. The variation of PiResNet-predicted relative error ($\varepsilon_a$) on the testing flows with the increasing noise level ($\delta$) in input features. Three activators are compared with the same regularization constraints both on network parameters and closure coefficients.

As aforementioned, the robustness of PiResNet stems from two parts: (i) the noisy-sensitivity of strain-dependent closure coefficients depends on network parameters and (ii) the noisy-sensitivity of learned model to integrity bases depends on the closure coefficients. The neuron-averaged value of 0.025 is constrained for network parameters, thus obtaining noisy-sensitive closure coefficients. Then we examine the closure coefficients, which is shown in Table III and Fig. 18. The closure coefficients of GEP are considerably large (details in Ref.[13]) and thus are omitted. Although complete integrity bases have been considered, our PiResNet with only first four terms can still reproduce the flow characteristics well for all testing cases. Weatheritt and Sandberg.[12] also found that more terms make limited gains in predictive accuracy. Thus, only four coefficients are shown. PiResNet provides a mean value of −0.077 for $c_1$ which is close to −0.09 in conventional models. As is the case in GEP, large values of closure coefficients in PiResNet appear on the duct bisector due to velocity gradients vanishing. The difference is that penalty on PiResNet-predicted closure coefficients (as shown in Eq. (18a)) helps regularize the solution of ill-posed problems and thus successfully obtain much smaller values than GEP. Coefficients ($c_1, c_2, c_3, c_4$) of PiResNet have small values below 0.18, which contributes to noisy-sensitivity to integrity bases ($\boldsymbol{s}, \boldsymbol{s}^2, \boldsymbol{s\omega} - \boldsymbol{\omega s}, \boldsymbol{\omega}^2$). Additionally, the small dispersion of closure coefficients, as shown in Table III, indicates good spatial smoothness. Larger values and a larger dispersion for closure coefficients occur when removing regularizations of closure coefficients. Obviously, regularization operations both on network parameters and closure coefficients do work well in improving the robustness of PiResNet.

TABLE III. The spatial average (Ave.) and root mean square (rms) of PiResNet-predicted closure coefficients for all testing flows.

|       | $c_1$  | $c_2$ | $c_3$  | $c_4$  |
|-------|--------|-------|--------|--------|
| Ave.  | −0.077 | 0.036 | −0.050 | −0.024 |
| rms   | 0.041  | 0.035 | 0.041  | 0.042  |

### D. Interpretability of closure coefficients

Finally, we analyze the physical roles of learned closure coefficients and their contribution to the production of turbulent kinetic energy $P_k \equiv -\boldsymbol{\tau}:\boldsymbol{S}$. $c_1<0$ in PiResNet is useful to numerical stability while $c_1>0$ in GEP as the wall is approached. $c_2>0$ in PiResNet indicate that $\boldsymbol{s}^2$ play an opposite role with the linear term $\boldsymbol{s}$, while $\boldsymbol{s\omega} - \boldsymbol{\omega s}$ with $c_3<0$ plays the same role with $\boldsymbol{s}$. $c_4<0$ in most regions and $c_4>0$ in the narrow corner region. Accordingly, negative $a_{ij}$ is dominated by $c_1$ while positive $a_{ij}$ is mainly controlled by $c_2$. The well-known in-plane secondary motion is almost accounted for by $c_2$, $c_3$ and $c_4$ due to $a_{22} - a_{33} \approx (c_2 + 2c_3 - c_4)(s_{12}^2 - s_{13}^2)/2$, which drives high-energy fluid from the near-wall region to the duct core. Notably, the symmetric distribution of ($c_1, c_2, c_3, c_4$) about the duct bisector at $Re_\tau=360$ with AR=1 is a physical reflection of embedded invariances in PiResNet. Furthermore, Eq. (14) in incompressible flows yields the PiResNet-predicted $P_k$ (nondimensionalized by $\varepsilon$) as

$$P_k/\varepsilon = -2c_1 \mathrm{tr}(\boldsymbol{s}^2) - 2c_2 \mathrm{tr}(\boldsymbol{s}^3) - 2c_4 \mathrm{tr}(\boldsymbol{\omega}^2 \boldsymbol{s}). \tag{23}$$

The disappearance of $c_3$ in Eq. (23) but its existence in Eq. (14) implies that $\boldsymbol{s\omega} - \boldsymbol{\omega s}$ plays a redistribution role between stress components and has no contribution to $P_k$, which is similar with the mechanism of pressure-strain correlations in transport equations. In duct flows, $\mathrm{tr}(\boldsymbol{s}^2) > 0$, $\mathrm{tr}(\boldsymbol{s}^3) < 0$, $\mathrm{tr}(\boldsymbol{\omega}^2 \boldsymbol{s}) > 0$ (negative in the corner region) and $\mathrm{tr}(\boldsymbol{s}^2) > |\mathrm{tr}(\boldsymbol{s}^3)| > |\mathrm{tr}(\boldsymbol{\omega}^2 \boldsymbol{s})|$. Thus, $\boldsymbol{s}, \boldsymbol{s}^2$ and $\boldsymbol{\omega}^2$ in PiResNet accelerate the production of turbulent kinetic energy while the linear term $\boldsymbol{s}$ with $c_1>0$ in GEP as the wall is approached, acts to hinder the production of turbulent kinetic energy. This behavior of GEP leads to a laminarization or "lift off" of the secondary structures.[13] To achieve physical and converged results, Weatheritt and Sandberg[13] claimed that an empirical modification of GEP-predicted $P_k$ is a must. Our PiResNet provides a correct mechanism for $P_k$. Accordingly, closure coefficients in PiResNet have clear groundings in physical arguments.

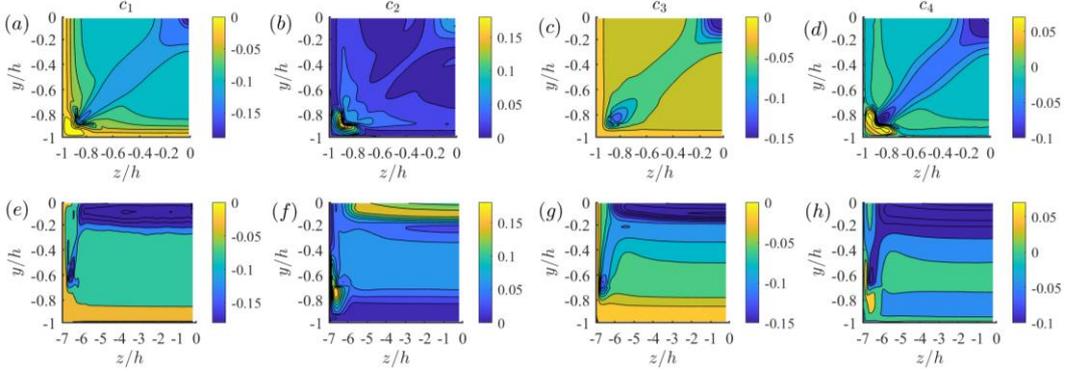

FIG. 18. PiResNet-predicted closure coefficients for a duct flow at (a)-(d) $Re_\tau$=360 with AR=1 and (e)-(h) $Re_\tau$=180 with AR=7. The first to forth columns corresponds to $c_1$, $c_2$, $c_3$ and $c_4$, respectively.

Overall, PiResNet is a successful model with closure coefficients of clear physical meaning, which provides realizable and robust predictions and can account for the variability of flow characteristics due to Reynolds number and aspect ratio.

## V. Summary and Perspectives

The current work has developed a universal turbulence modeling framework under which an invariant, realizable, unbiased, and robust data-driven turbulence model was achieved. At every phase of the model development lifecycle, the underlying physical considerations are indispensable: domain-knowledge has been attentively inferred and reasonably converted to modeling knowledge embedded both in the design and training processes. In such endeavors, the methodology addressed in this work has been insured as an interpretable machine learning framework for turbulence modeling.

In the design phase: first and foremost, the objectivity of the proposed PiResNet was strictly preserved by suitably choosing extended Galilean-invariant input features to regress the closure coefficients. This principle prevents from a preference for the coordinate system adopted by the modeler, which is user-friendly. Second, a dimensionless quantity ($Re_T$) was introduced to parameterize the newly-proposed turbulent time scale. As an additional argument, it successfully helped extricate from possible nonunique mappings of conventional inputs (only dimensionless strain magnitude and vorticity magnitude for coefficient regression) to outputs of a RANS model, thus being possible to employ machine learning algorithms to establish a deterministic functional relation. Third, realizability as one of the contents of turbulence physics was reasonably respected to constrain the well-learned PiResNet, especially when the model is operating in an extrapolatory mode. Using the PIR corrector, realizable solutions were gained in the predictive settings. It is worth noting that $Re_T$ and the PIR corrector can also be useful to other algebraic RANS models.

In the training phase: a fair learning strategy comprising of unbiased data sampling (here referring to the clustering-based resampling technique) and unbiased objective-function design was successfully implemented to update internal network parameters. The unbiased objective-function design contains two-fold: (i) a scale- and frame-invariant objective-function to insure the hyper-parameter settings of PiResNet against being invalid across training datasets of dynamically similar flows only with different scales or different coordinate systems, which is useful for other modelers to reproduce this work; and (ii) effective measures taken for the first time to rebalance more fairly each contribution of the anisotropy components to the overall training error. Besides, regularizations were performed both on trainable network parameters and regressed closure coefficients to reduce the PiResNet sensitivity to the uncertainty of input data, thus achieving a robust model. The former accounts for the sensitivity of the closure coefficients to the uncertainty of invariant inputs, while the latter for the amplifier of the uncertainty of tensor bases.

It cannot be overstated that physical considerations are critical to data-driven turbulence modeling. The above-mentioned measures jointly do improve the predictive performance of data-driven turbulence modeling based on limited data and are significant efforts towards its practical use in the engineering environment in the near future. Also, the noisy-insensitive PiResNet can provide high-fidelity predictions of all Reynolds-tress components based on sparse experimental data. Based on the PiResNet-predicted Reynolds-stress and the RANS solver, one can easily develop a hybrid experiment-CFD method. Despite encouraging results, on the other hand, there is much to be gained by further calibrating the RANS-related transport equations, enriching the diversity of training data, and carefully exploring a broader set of input features.

We do not claim that the learned PiResNet necessarily is applicable to other untested complex flows although it does gain good generalization across some two-dimensional and three-dimensional flows. PiResNet needs to be widely validated. Rather, the philosophy and formalisms employed in the UIML are of a general nature, and not restricted to the type of RANS-based closure, the type of neural netwoork architecture, or the type of training case. Following the guidelines of UIML, one can easily extend to retrain a data-driven model, even at LES to predict more complex flow configuraions.

However, the most challenging issue remains how to develop a generalizable model across multiple classes of complex flows. Therefore, building the bechmark dataset containing diverse typical flow phenomenon (e.g., secondary effects, flow separation, streamline curvature, to name a few) is presently eager to develop and evaluate data-driven turbulence models. Additionally, a complete and compact set of input features should be investgated sysmatically. The incompleteness and redundancy may destroy the generalization capability. Most importantly, data-driven turbulence modeling to date (including PiResNet) has been mainly centerd around machine-learning-augmentd calibration of the closure coefficients given a conventional form-constraint RANS model, which only improves the parametric inadequacy. It is common knowledge that the inadequacy in model-form is the crux. Towards this end, it is indeed essential to develop a "form-free-in-prior" data-driven model to directly infer the model-form in adequately representing the rich dynamics of turbulence. This requires that machine learning algorithms can directly deal with tensor problems, at least vector problems (when using a spectral decomposition). Future work in this direction may be of great interest.